\documentclass[prb]{revtex4-1}

\usepackage{graphicx}

\begin{document}



\title{Planetary Atmospheres as Non-Equilibrium Condensed Matter}
\author{J. B. Marston}
\affiliation{Department of Physics, California Institute of Technology, MC114-36, Pasadena, CA 91125 USA 
and Department of Physics, Box 1843, Brown University, Providence, RI 02912-1843 USA}


\begin{abstract}
Planetary atmospheres, and models of them, are discussed from the viewpoint of condensed matter physics.  Atmospheres are a form of condensed matter, and many  interesting phenomena of condensed matter systems are realized by them.  The essential physics of the general circulation is illustrated with idealized 2-layer and 1-layer models of the atmosphere.  Equilibrium and non-equilibrium statistical mechanics are used to directly ascertain the statistics of these models. 
\end{abstract}

\maketitle

\section{Introduction}
\label{introduction}

Planetary atmospheres exhibit many of the most interesting aspects of condensed matter systems.  Surprising collective behaviors emerge from the interaction of many degrees of freedom and include phenomena such as broken symmetry, dimensional reduction,  condensation, phase transitions, scaling, and the dynamical protection of conserved quantities.  Theoretical descriptions of atmospheres include such familiar themes as hierarchies of complexity, field versus particle bases, correlations, perturbative versus non-perturbative approximations, linear response, and transport.   Even quantum many-body theory makes an appearance.  Though atmospheric dynamics are purely classical, a formalism that resembles the many-body Schr\"odinger equation describes the stationary statistics.   Such direct statistical descriptions are still in their infancy, and require further testing and development, but have the potential to greatly improve our understanding of climate.  Work along these lines, described below, lays the foundations for a new class of climate models that could be both more accurate and computationally efficient than present-day models that run on the largest supercomputers.  

This review complements and expands upon a short article that appeared in {\it Physics Trends}\cite{Marston:2011p604}.  I will illustrate the various aspects mentioned above by discussing the observed general circulation of the atmosphere, and the behavior of prototypical and simplified, yet highly non-trivial, models of planetary atmospheres.  Such reduced models isolate the essential physics of the atmosphere, and enable the development of new direct statistical approaches.  No attempt is made to describe atmospheres or models of them in any detail; for that the reader is referred to standard textbooks and review articles\cite{Peixoto:1992,salmon98,Vallis:2006,Schneider2006}.  Laboratory-scale experiments, often with rotating tanks, play an important role in understanding atmospheric processes (see for instance Refs. \onlinecite{Huang:1994p503,Weeks:1997p477,CastrejonPita:2010p392,Shats:2010p545}) but the emphasis here will be on theoretical models.  The equations of motion will be cast in coordinate-independent form to highlight physics rather than the analytical and numerical tools needed to simulate them.  The focus is on large-scale features of planetary atmospheres that should appear at least somewhat familiar to condensed matter physicists.  Condensed matter physics also plays a more traditional role in understanding planetary structure, for instance in modeling the structure of water and hydrogen under extreme pressure\cite{Militzer:2010p621} but this connection will not be explored.  When condensed matter concepts first appear below they will be italicized to emphasize connections with atmosphere dynamics.

The time is ripe for a look at planetary atmospheres from the perspective of condensed matter physics. Besides obvious and growing concern with climate change, a plethora of data from exoplanets (planets orbiting other stars) of unanticipated diversity is now being accumulated, including first observations of the atmospheres.  Reconstructions of past Earth climates are also advancing, and understanding both exoplanets and paleoclimates will require new tools.  It is hoped that this overview will spur the cross-disciplinary fertilization of ideas. 

The outline of the article is as follows.  Section \ref{components} presents an overview of the climate system, and its component subsystems.   The hierarchy of models and two prototypical atmospheric models of reduced complexity are presented in Section \ref{reducedModels}.  These are used in the remainder of the article to illustrate key notions of atmospheres with an eye to their condensed matter physics.  Section \ref{nearEquilibrium} discusses work by a number of researchers on statistical solutions of single-layer models that are close to equilibrium.  By virtue of being near equilibrium, some nearly exact conclusions may be drawn regarding the model statistics.   Most interesting atmospheric processes are far from equilibrium, however, and these require new approaches for their statistical description.  One such approach, based upon an expansion in cumulants, is presented in Section \ref{nonEquilibrium}.  Two alternative approaches are presented in Section \ref{otherApproaches}.  Finally a brief summary and some open questions are presented in Section \ref{conclusion}.  

\section{Components of Climate}
\label{components}
 
The climate system of Earth is complicated with many strongly interacting degrees of freedom.  Most efforts to understand it begin by disassembling the system into component subsystems, with the expectation that the components, if sufficiently simple by themselves, will be amenable to detailed analysis.   The problem is a familiar one to any physicist who works on complex systems.  Condensed matter physics for instance offers many such examples, and a generally accepted procedure for modeling such systems is to begin with highly simplified models, adding to their complexity only as the requirements of the physics dictate.  Isaac Held has advocated taking the same approach to climate\cite{Held2005}.  

\begin{figure}
\centerline{\includegraphics[width=4in]{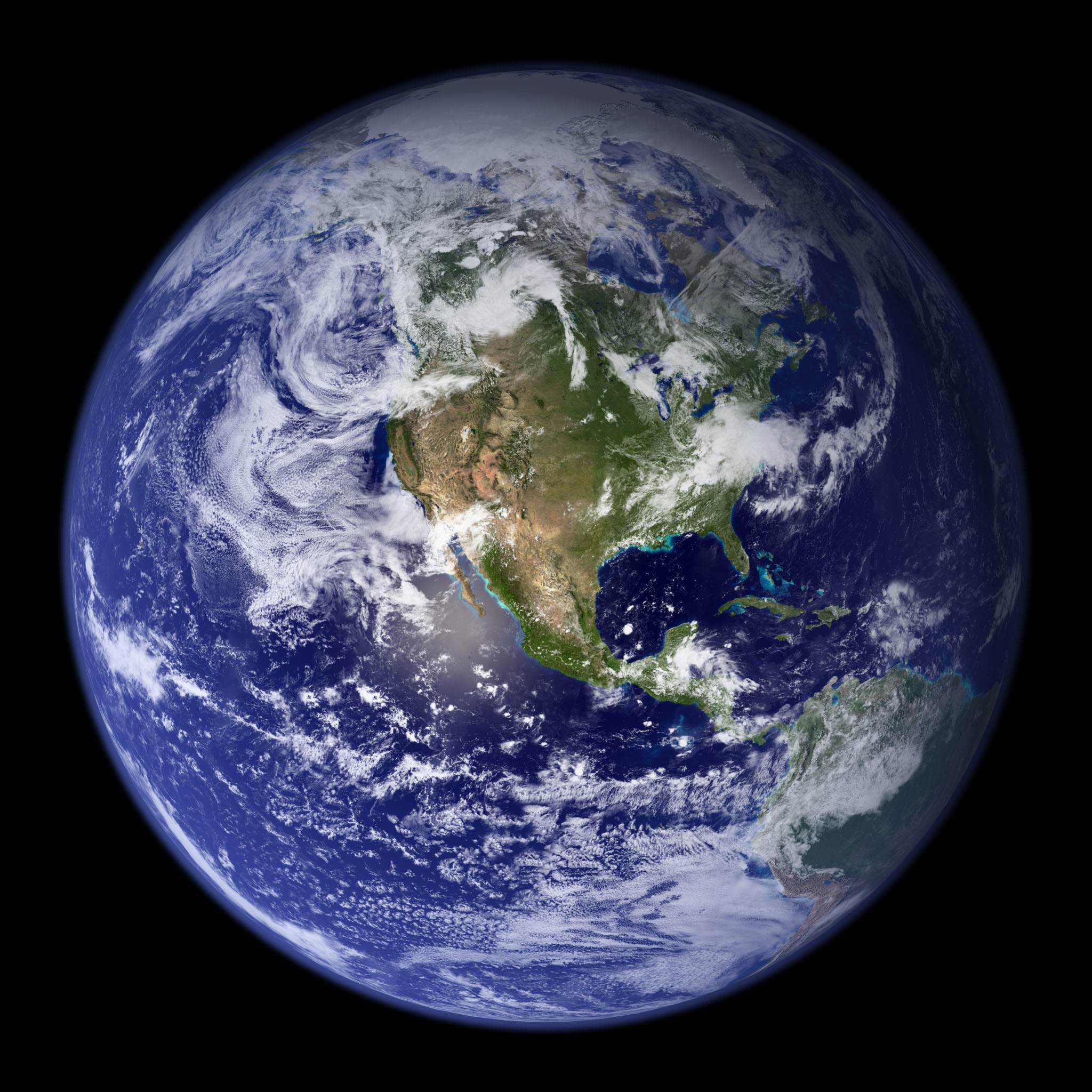}}
\caption{\label{figure1} View of Earth from space showing the major component of the climate system: the atmosphere, biosphere, cryosphere, hydrosphere, and lithosphere.  Credit: NASA Earth Observatory.}
\end{figure}

Following this approach, the climate system may be divided into the biosphere (life), lithosphere (land surfaces), hydrosphere (oceans), atmosphere, and cryosphere (snow and ice)\cite{Peixoto:1992}.  Each of these components is visible in a photograph of the earth from space (Fig. \ref{figure1}).  Other planets in the solar system are missing some of the components, though possibly some exoplanets (planets outside of the solar system) will be found to possess all of them.  Earth Systems Models (ESMs) running on supercomputers attempt to simulate all of the component subsystems together.   However it can be difficult to isolate and understand key processes in such comprehensive models. 

On long time scales, the carbon cycle -- the flow of carbon between reservoirs such as the air, oceans, and biosphere --  responds to climate, creating feedbacks that act back upon the climate.  Ice core records show that the concentrations of atmospheric carbon dioxide and methane are closely coupled to global temperatures, but the mechanisms and timescales of the couplings are unclear.  Changes in vegetation also affect the amount of light that land surfaces reflect back out into space (the fraction known as the ``albedo'') as well as the transpiration of water into the air.  A well-understood framework captures the mathematics of feedbacks\cite{Roe:2009p471} but quantitative models for many feedback processes remain primitive with many uncertainties.  

The dynamics of oceans and ice sheets also operate on time scales that are long compared to most atmospheric processes\cite{salmon98,Vaughan:2007p218}.  A statistical understanding of the atmosphere that would amount to ``integrating out'' the fast atmospheric degrees of freedom, yielding a more tractable climate system in which the remaining slow modes in the hydrosphere, cryosphere, lithosphere and biosphere) would still be modeled explicitly would be very desirable.  If this goal could be achieved, it would enable simulations of much longer time scales, such as the ``deep time'' studies of paleoclimates.  

This review focuses only on the subsystem with the fastest evolving degrees of freedom, the atmosphere.   As a first approximation the other subsystems are ignored, except in so far as they provide boundary conditions for the atmosphere.  Comprehensive models of climate must of course simulate all of the subsystems, but the goal here is to instead isolate different sources of complexity within the atmosphere.   There are many fundamental questions that remain unanswered about atmosphere dynamics, including the following\cite{Stone:2008p639}:  What is the magnitude of the heat transported poleward by the atmosphere?  How does small-scale convection compare in importance to large-scale eddies?  And what dynamical processes contribute to the statistical steady state of the atmosphere?

\subsection{Atmospheres}
\label{atmospheres}

The atmosphere is, to a good approximation, governed by a set of seven equations.  The first three are the equations of motion for the three components of the momentum of a parcel of air.  Next there is the equation of continuity for the air that reflects the law of conservation of mass.  There is a separate equation for water vapor:  It is conserved by advection, but can also undergo phase transitions to water or ice, and enter or leave the atmosphere through several sources and sinks.  The sixth equation is the first law of thermodynamics (conservation of energy).  Finally there is the equation of state for air (including water vapor).  Standard approximations then made to these equations include filtering out sound waves (that carry little energy in atmospheres, and thus are an unnecessary complication).  

In principle the seven equations describe clouds as well as the motion of the air.   The amount of condensed water in the atmosphere is two orders of magnitude less than the amount of water in the vapor phase.  Yet 
clouds are vexing for models because they interfere with both incoming visible and outgoing infrared radiation 
with the nature of their interaction with the radiative transfer of energy varying with cloud type.  Low stratocumulus clouds in the marine boundary layer are particularly effective at cooling the planet as they reflect light and also radiate at a high temperature.  High cirrus clouds, by contrast, transmit visible light but radiate in the infrared at a much lower temperature and hence act to warm the atmosphere.  Clouds are also problematic because cloud processes operate at length scales ranging from nanometer-size aerosol particles to convection and turbulence at intermediate scales, to clustering of clouds at scales of hundreds of kilometers.  Indeed the largest uncertainty in climate models comes from limits to our understanding and modeling of clouds\cite{Stephens05}.  In the highly simplified models considered below I remove this complication by only considering dry atmospheres which still retain much of the interesting physics of mid-latitude dynamics.  

Geophysical fluids are often highly stratified, moving for the most part in the horizontal directions\cite{Xia:2011p598}.  Layered clouds, so ubiquitous outside of the tropics, make this stratification readily apparent.  This {\it dimensional reduction} is a consequence of planetary rotation and convection (the latter acts to restore stratification).    A commonly made approximation is to assume that acceleration in the vertical direction is negligible compared to acceleration due to pressure and gravity (hydrostatic equilibrium).  It is furthermore convenient to resolve the horizontal components of the velocity field into rotational and divergent parts\cite{Lorenz:1960p483,Heikes:1995p120}:
\begin{eqnarray}
\vec{v} = \hat{r} \times \vec{\nabla} \psi + \vec{\nabla} \chi
\end{eqnarray}
where $\hat{r}$ is the unit vector pointing in the radial direction, $\psi$ is the scalar streamfunction and $\delta \equiv \vec{\nabla} \cdot \vec{v} = \nabla^2 \chi$ is the scalar two-dimensional divergence.  The fluid is effectively incompressible because the wind speed is small compared to the speed of sound (the Mach number is much less than one) so $\delta \neq 0$ implies vertical motion.  The relative vorticity $\zeta$ is the curl of the velocity field and points in the vertical direction,
\begin{eqnarray}
\zeta \equiv \hat{r} \cdot (\vec{\nabla} \times \vec{v}) = \nabla^2 \psi\ .
\label{vorticity}
\end{eqnarray}
The absolute vorticity $q \equiv \zeta + f(\phi)$ is the vorticity as seen in an inertial frame where $f(\phi) = 2 \Omega \sin \phi$ is the Coriolis parameter or planetary vorticity familiar from the physics of Foucault's pendulum ($\phi$ is the latitude and $\Omega$ is the angular rotation rate of the planet).  The planetary vorticity is in some ways analogous to the effect of a magnetic field on the orbital motion of electrons in condensed matter and in plasmas.  Because the dimensionless Rossby number $R_o \equiv |\zeta / f | \ll 1$ away from the tropics, the Coriolis force is dominant and wind motion is close to ``geostrophic'' meaning that the wind blows nearly at right angles to the pressure gradient.  Typically also in the extratropics $|\zeta| \gg |\delta|$ so vorticity is the more important of the two variables describing flows there.

\subsection{Parameterizations}
As it stands the equations of motion are still too difficult to solve numerically, because they describe processes that occur on length scales that are small compared to the grid scale of a global numerical simulation.  In practice these processes are parameterized, meaning that simplified and semi-phenomenological models are used to capture them instead of direct simulation.  Examples include: Land-surface atmosphere parameterizations, water-atmosphere parameterizations, planetary boundary layer and turbulence parameterizations, convective parameterizations, cloud microphysics parameterizations, radiation parameterizations, cloud cover-radiation parameterizations, and parameterizations of drag due to topography\cite{stensrud:2007}.

Parameterizations can be complex or simple.  In the following the simplest possible parameterizations are adopted that still retain the essential physics.  For instance the coupling of radiation to the atmosphere\cite{Pierrehumbert:2009p206} will be described phenomenologically by Newtonian relaxation to a prescribed temperature profile\cite{Held:1994p409}; friction in the planetary boundary layer is parameterized by a Rayleigh friction constant.  

\section{Hierarchy of Atmospheric Models of Reduced Complexity}
\label{reducedModels}

The Hubbard model is a highly simplified model of condensed matter physics that highlight the key physics of metal-insulatior transitions, antiferromagnetism, and other phenomena while at the same time suppressing details of real materials such as extra bands or long-range interactions that would distract from a deeper understanding of the phenomena.  Likewise models of reduced complexity function as prototypes of the large-scale circulation of atmospheres.  Two such deterministic models are described below. The simplicity of the models makes possible a near-complete understanding of their rich behavior.  However, just as first-principles ab-initio methods such as density functional theory are needed for the quantitative modeling of materials, detailed modeling of real climate requires models of full complexity and realism.   Understanding attained from simplified models informs work with such comprehensive models.

The two atmospheric models of reduced complexity are studied below and in subsequent sections by direct numerical simulation (DNS) and by statistical approaches.  The traditional way to accumulate statistics is by sampling DNS at regular time intervals.  As an alternative, it is possible to solve directly for the statistics.  Such direct statistical simulation (DSS) is still in its infancy, and the various approaches and approximations require testing against statistics obtained by traditional DNS.  As discussed in Section \ref{nonEquilibrium} DSS offers potential advantages over DNS and may in time be applied to models of increasing complexity and realism.

\subsection{Two Layer Model}
\label{two-layers}

\begin{figure}
\centerline{\includegraphics[width=5in]{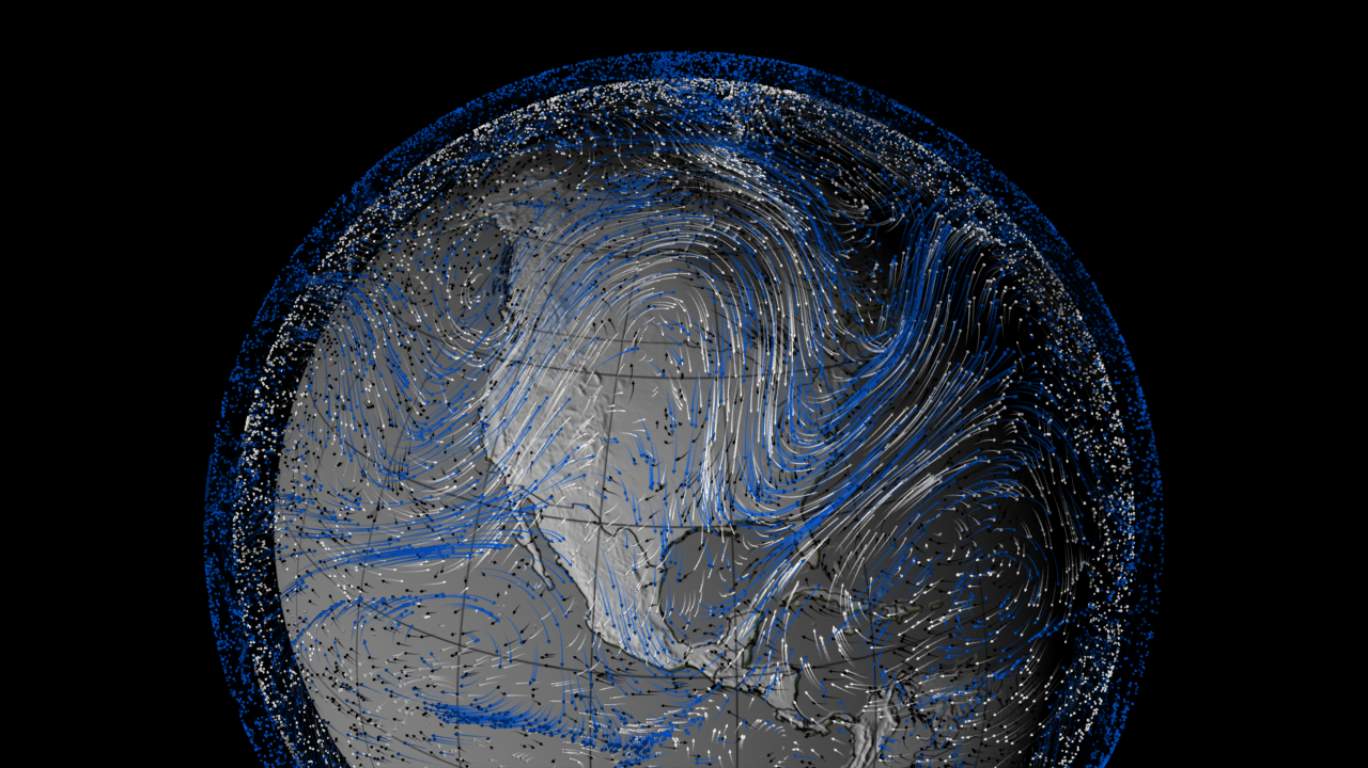}}
\vskip 0.5cm
\centerline{\includegraphics[width=5in]{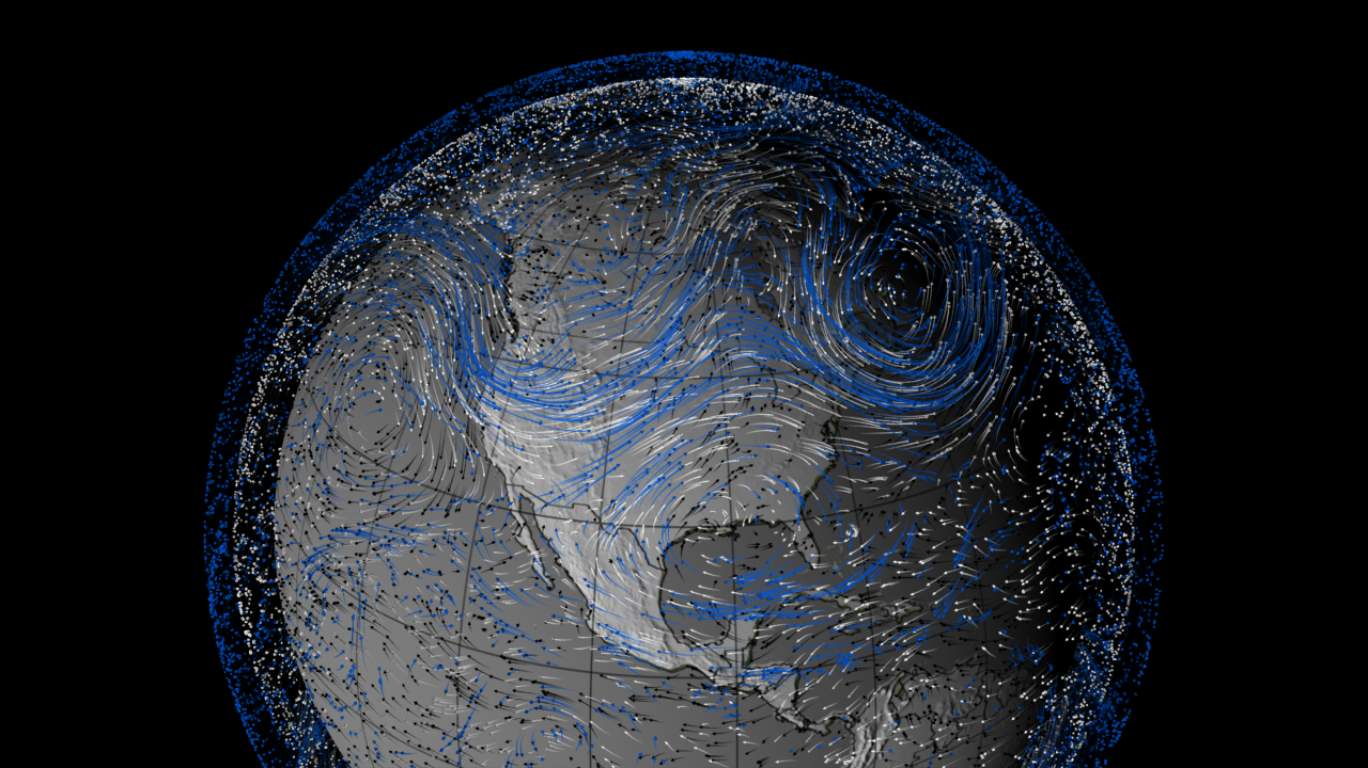}}
\caption{\label{figure2} Wind velocity fields at three different heights (black arrows at $1.5$ km; grey at $5.4$ km; and blue at $9.2$ km) reconstructed for 1988 (top) and 1993 (bottom).  Evident in both figures is the mid-latitude zonal jet that flows from west to east.  By contrast in the tropics the trade winds flow from east to west.  Vertical as well as horizontal shear is apparent as wind speeds generally increase with altitude.  High pressure over the United States in the summer of 1988, visible as geostrophic flow moving clockwise around the high plains, led to a prolonged drought.  Credit: NASA / Goddard Space Flight Center Visualization Studio and NASA Earth Observatory (October 5, 2010).}
\end{figure}

As a major simplification the atmosphere is modeled by only two layers in the vertical direction labelled as $0$ and $1$ and located respectively at heights with corresponding pressures that are $3/4$ and $1/4$ of the surface pressure\cite{Lorenz:1960p483,Held:1978p173,Lee:1993p461,Ringler:2000p118}.  Differential heating of the two layers drives atmospheric motion. (Figure \ref{figure2} is a visualization of actual motion in Earth's atmosphere at three different heights.)  Also, as mentioned above, all water is removed for the sake of simplicity.   Thermodynamics in this dry atmosphere is then most simply described in terms of the potential temperature $\theta$ which is the temperature that a parcel of air would have if it were moved adiabatically to a reference pressure of one atmosphere.    

The advection of air in the atmosphere, and the transport of heat along with it, may be nicely expressed in a coordinate-independent way by introducing two scalar bilinear differential operators\cite{Ringler:2000p118}, the Jacobian $J[,~ ]$ and flux-divergence $F[,~ ]$:
\begin{eqnarray}
J[A,~ B] &\equiv& \hat{r} \cdot (\vec{\nabla} A \times \vec{\nabla} B)
\nonumber \\
F[A,~ B] &\equiv& \vec{\nabla} \cdot (A \vec{\nabla} B)\ 
\end{eqnarray}
where the derivatives act in the horizontal directions only. 
It is also convenient to decompose fields in the two layers into symmetric or barotropic $\overline{A} \equiv \frac{1}{2} (A_1 + A_0)$ and antisymmetric or baroclinic $\widehat{A} \equiv \frac{1}{2} (A_1 - A_0)$ modes.  Thus the barotropic and baroclinic components of the bilinear field $A B$ are given by:
\begin{eqnarray}
\overline{A B} &=& \overline{A}~ \overline{B} + \widehat{A}~ \widehat{B}
\nonumber \\
\widehat{A B} &=& \widehat{A}~ \overline{B} + \overline{A}~ \widehat{B}\ .
\end{eqnarray}

The barotropic divergence $\overline{\delta} = 0$ in the model because any air leaving one layer enters the other.  In the absence of any forcing or damping the equations of motion for the scalar fields in the two layers are:
\begin{eqnarray}
\dot{\overline{q}} &=& J[\overline{q},~ \overline{\psi}] + J[\widehat{q},~ \widehat{\psi}] 
- F[\widehat{q},~ \widehat{\chi}] - J[\widehat{\delta},~ \widehat{\chi}] - F[\widehat{\delta},~ \widehat{\psi}]
\nonumber \\
\dot{\widehat{q}} &=& J[\widehat{q},~ \overline{\psi}] + J[\overline{q},~ \widehat{\psi}] 
- F[\overline{q},~ \widehat{\chi}] 
\nonumber \\
\dot{\widehat{\delta}} &=& J[\overline{q},~ \widehat{\chi}] + F[\widehat{q},~ \overline{\psi}] 
+ F[\overline{q},~ \widehat{\psi}] - \nabla^2 (\widehat{K} + C_p B \overline{\theta})
\nonumber \\
\dot{\overline{\theta}} &=& J[\overline{\theta},~ \overline{\psi}] + J[\widehat{\theta},~ \widehat{\psi}] 
- F(\widehat{\theta}, \widehat{\chi})
\nonumber \\
\dot{\widehat{\theta}} &=& J[\widehat{\theta},~ \overline{\psi}] + J[\overline{\theta},~ \widehat{\psi}] 
- F(\overline{\theta}, \widehat{\chi}) + \overline{\theta} \widehat{\delta}
\label{two-layer}
\end{eqnarray}
The first three equations describe the atmospheric flow in the two layers, driven by differences in temperature, and hence pressure, and deflected by the Coriolis force.  The last two equations for the mean potential temperature $\overline{\theta}$ and static stability $\widehat{\theta}$ describe the transport of heat by the flow.  
$K(\psi, \chi) = \frac{1}{2} [F(\psi, \psi) - \psi \nabla^2 \psi + F(\chi, \chi) - \chi \nabla^2 \chi] + J[\psi, \chi]$ is the kinetic energy density per unit mass density, $C_p$ is the (constant pressure) specific heat of dry air and 
$B \equiv \frac{1}{2 a^2}~ [ (\frac{3}{4})^{R/C_p} - (\frac{1}{4})^{R/C_p}]$ where the adiabatic exponent $R/C_p = \frac{2}{7}$ for ideal diatomic gases, and $a$ is the planetary radius.   The equations of motion, Eq. \ref{two-layer}, are quadratically nonlinear.  As it stands they conserve total angular momentum, mean potential temperature, circulation, total energy 
\begin{eqnarray}
E = \int \bigg\{ \overline{K} + \widehat{K} - C_p B~ \widehat{\theta} \bigg\}~ d^2\Omega~ ,
\end{eqnarray}
and the mean square potential temperature.  

The simplest way to model the effects of solar radiation and long-wave emission is by relaxing the potential temperature towards a prescribed meridional profile.  To the two equations in Eqs. \ref{two-layer} for $\theta$ the following terms are added:
\begin{eqnarray}
\dot{\overline{\theta}} &=& \ldots + (\overline{\theta}_0 - \overline{\theta}) / \tau_R
\nonumber \\
\dot{\widehat{\theta}} &=& \ldots + (\widehat{\theta}_0 - \widehat{\theta}) / \tau_R
\end{eqnarray}
where $\tau_R \approx 30$ days is a typical timescale for the radiative heating and cooling of the atmosphere.  Prescribed functions of latitude $\overline{\theta}_0(\phi) = \overline{\theta}_0 \times (\frac{1}{3} - \sin^2 \phi)$ and  $\widehat{\theta}_0(\phi) = \widehat{\theta}_0 \times \cos^2 \phi$ approximate the radiative driving\cite{Held:1994p409}.  Energy so input is then removed through frictional coupling of the lower layer to the surface, parameterized most simply by adding damping terms to the first three mechanical equations of motion in Eq. \ref{two-layer}:
$\dot{\overline{q}} = \dots - \kappa (\overline{q} - \widehat{q})$, $\dot{\widehat{q}} = \dots - \kappa (\widehat{q} - \overline{q})$, and 
$\dot{\widehat{\delta}} = \dots - \kappa \widehat{\delta}$.  Because the dimensionless Reynolds number $R_e \equiv V L / \nu$ is extremely large for large-scale atmospheric motion, viscous dissipation can be ignored.  

\subsection{Single Layer Model}
\label{single-layer}

A further simplification is possible, the reduction of the two layers to just one.  
In this barotropic limit pressure is a function only of density, and not of temperature, and there is no variation of the fluid motion with height.  Setting $\overline{\theta} = const$ and 
$\widehat{\theta} = 0$ in Eqs. \ref{two-layer} yields height-independent solutions for which the baroclinic modes have no amplitude, $\widehat{\delta} = \widehat{q} = 0$.  The fluid then satisfies the single-layer barotropic vorticity equation: 
\begin{eqnarray}
\dot{q} = J[q, \psi] 
\label{barotropic}
\end{eqnarray}
or equivalently
\begin{eqnarray}
\frac{D q}{D t} \equiv \frac{d q}{d t} + \vec{v} \cdot \vec{\nabla} q = 0
\label{quasigeostrophic}
\end{eqnarray}
with $\overline{q}$ now written as simply $q$.   As it stands this equation means that absolute vorticity is conserved under advection.  Various linear forcing and damping terms can then be added to the right hand side of Eq. \ref{quasigeostrophic} to enable the investigation of various simple fluid models such as those described in Section \ref{nearEquilibrium} and Subsection \ref{Jet} below.

\section{Near Equilibrium Fluids}
\label{nearEquilibrium}

\begin{figure}
\centerline{\includegraphics[width=2.4in]{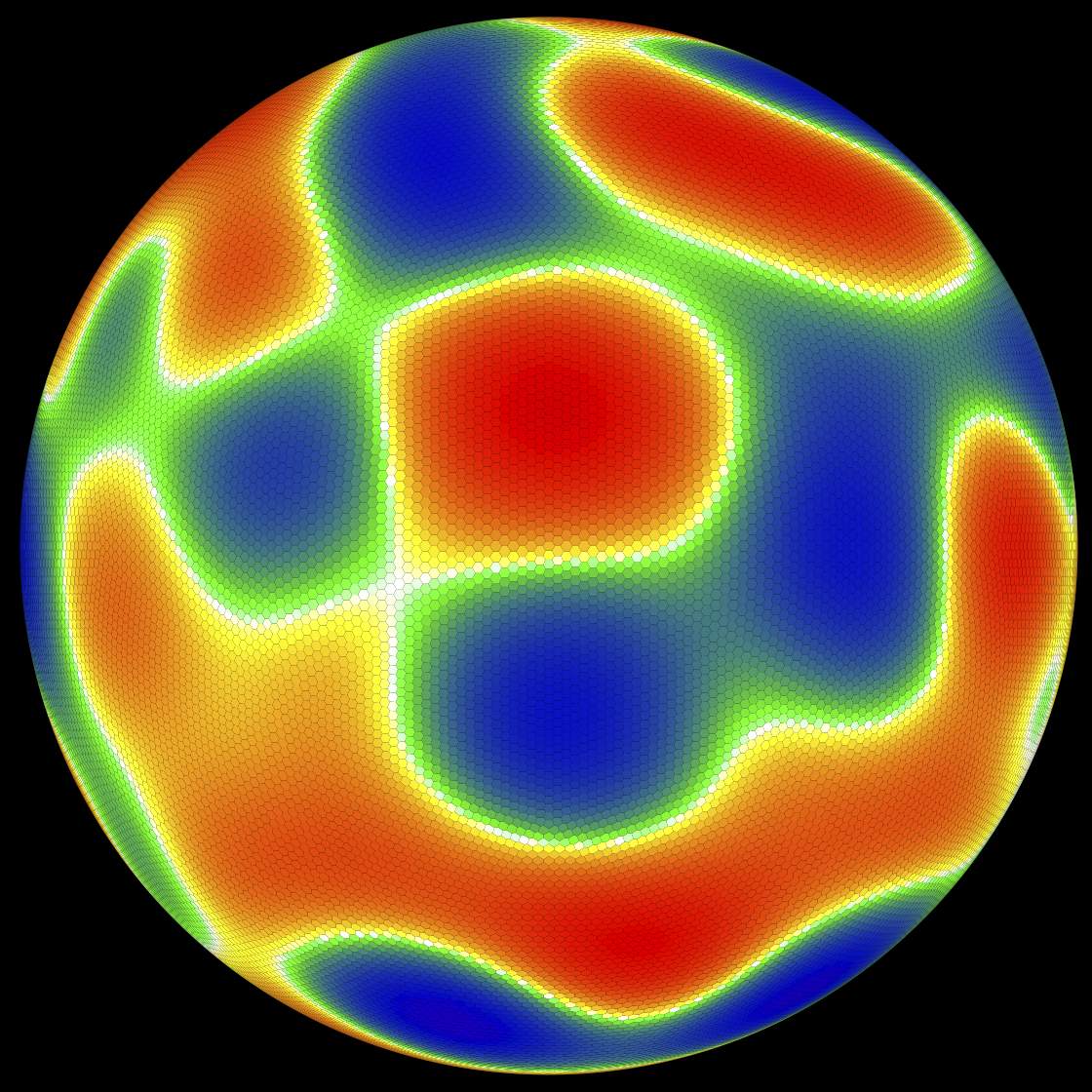} \includegraphics[width=2.4in]{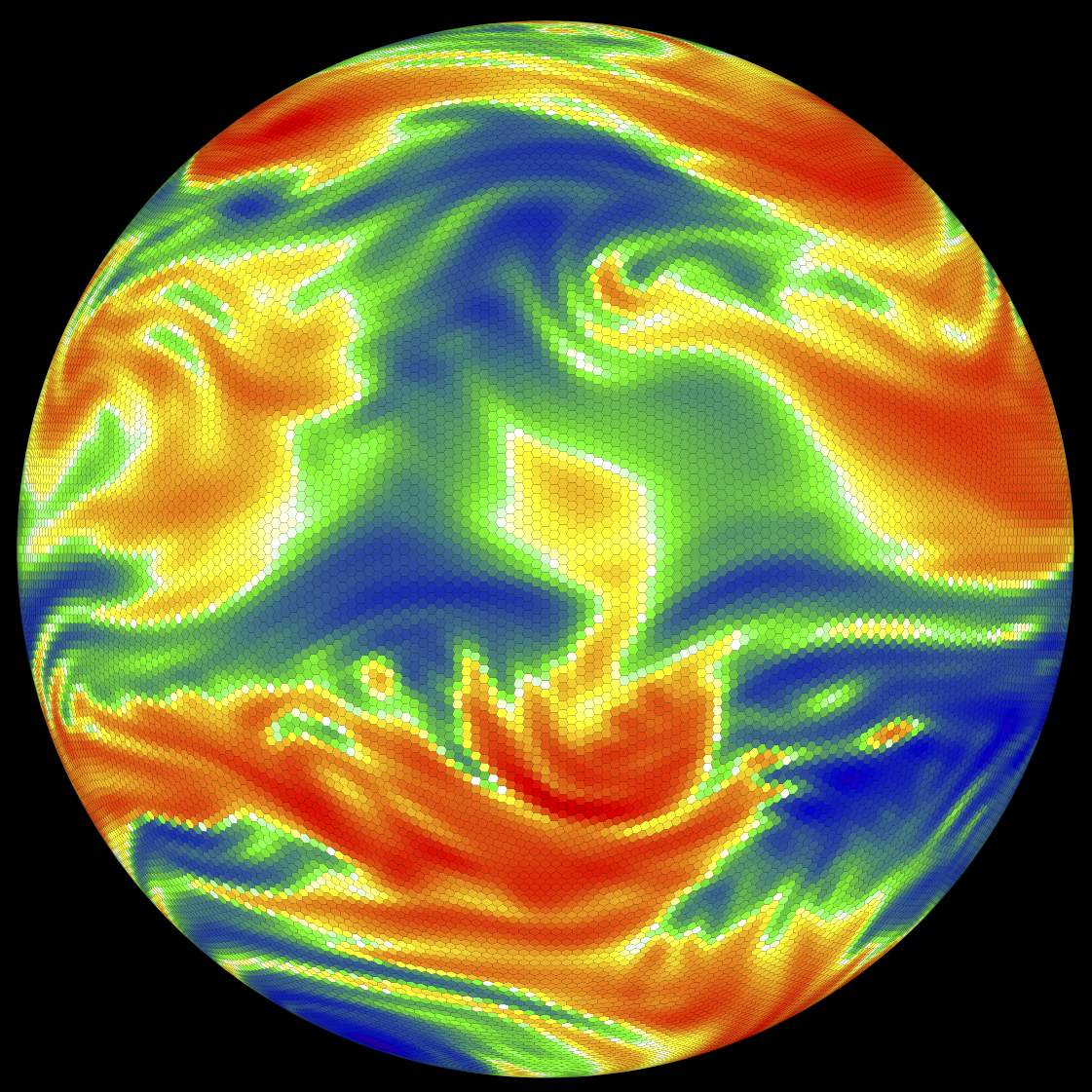}}
\vskip 0.1cm
\centerline{\includegraphics[width=2.4in]{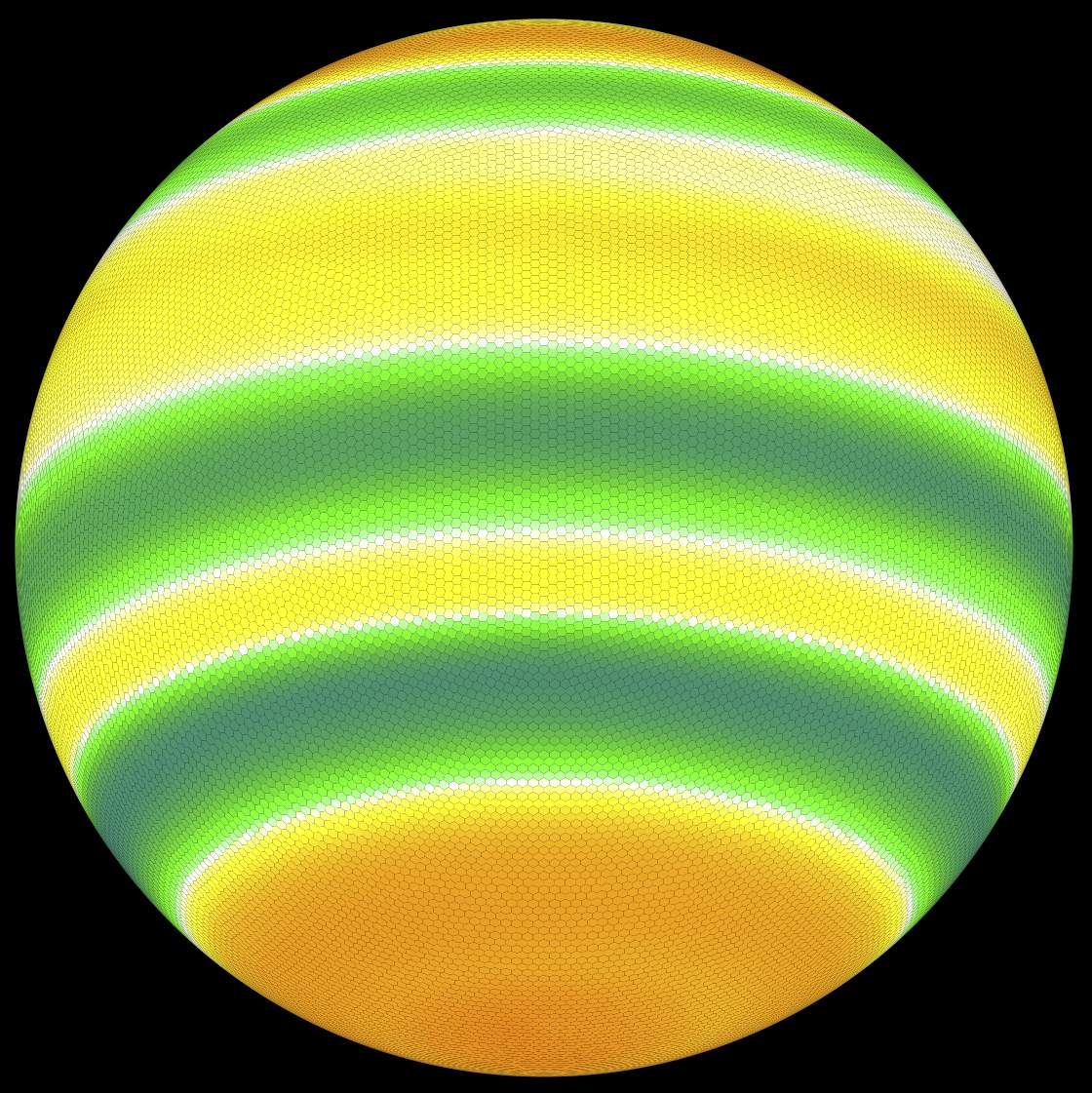} \includegraphics[width=2.4in]{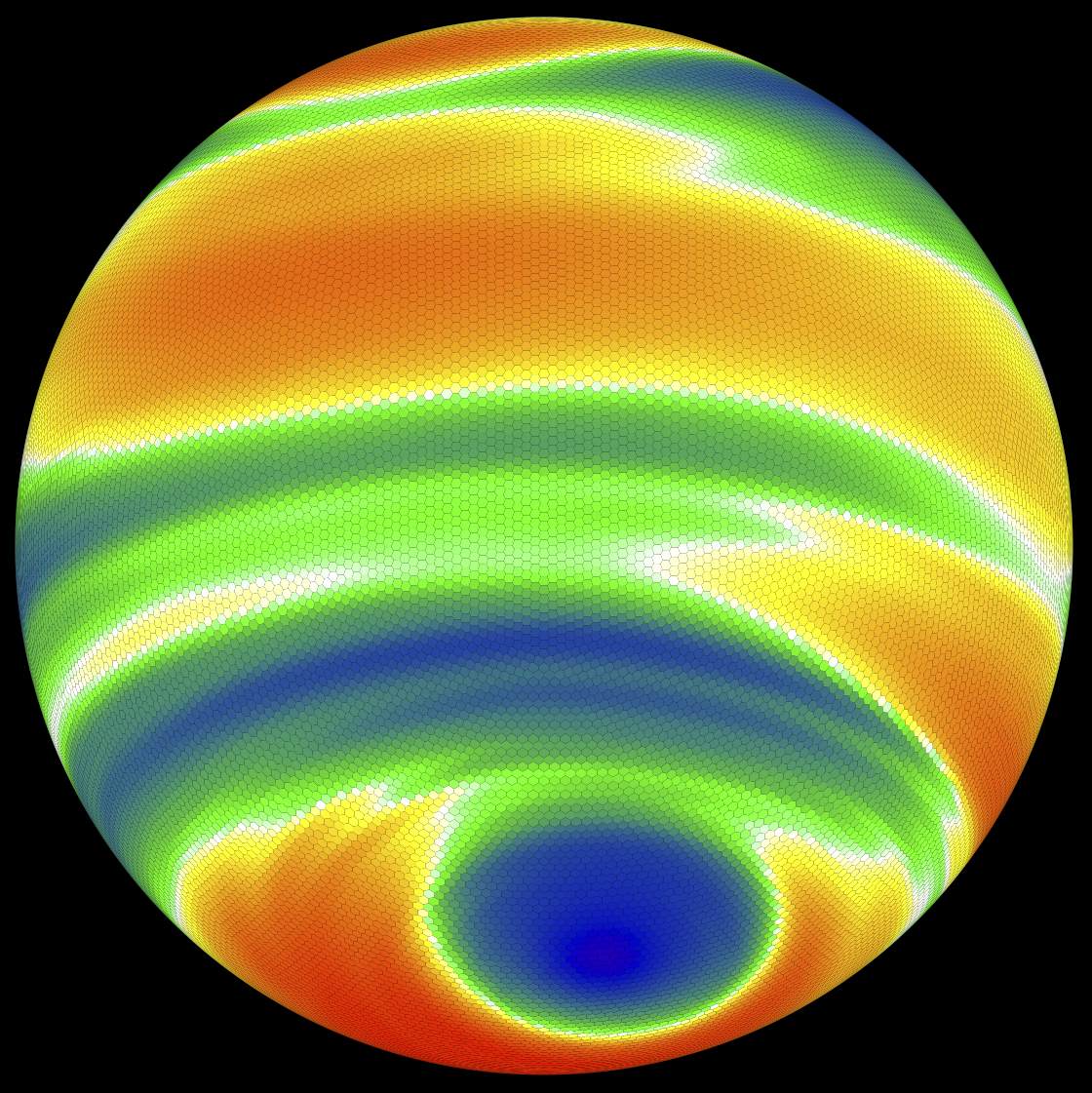}}
\vskip 0.1 cm
\centerline{ \includegraphics[width=4.85in]{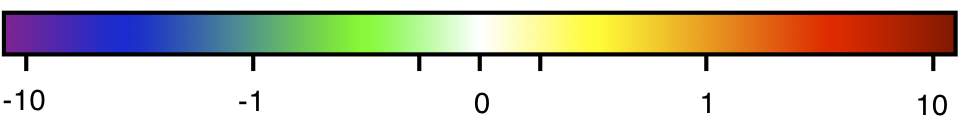}}
\caption{\label{figure3} Time evolution of unforced relative vorticity on a sphere rotating once per day, damped only by hyperviscosity, as viewed from a point directly above a latitude of $15^\circ$ S.  The eddy turnover time is approximately $1$ day.  Clockwise from upper left: $t = 0$ days, $33.7$ days, $260.8$ days, and time-averaged vorticity.  The flow first develops fine structures due to nonlinear advection.  It evolves at long times into a mixture of zonal jets and an isolated vortex in the southern hemisphere.   The time-averaged vorticity shows the expected recovery of azimuthal symmetry in the statistic.  The direct numerical simulation here is performed on a spherical geodesic grid\cite{Randall:2002p119} with $40,962$ cells.  In the figure, and those that follow, the unit of time is 1 day, and the unit of distance is 1 planetary radius.}
\end{figure}

Two-dimensional turbulent flows organize spontaneously into large-scale and long-lived coherent structures, such as those seen in Fig. \ref{figure3}.   In this DNS calculation Eq. \ref{barotropic} on the unit sphere has been integrated forward in time starting from an initial condition with a random pattern of vorticity (upper left panel) until only zonal jet streams, signaled by bands of vorticity of alternating sign, and a single vortex remain.    The coherent structures that {\it condense} out of such freely decaying turbulence are different than those generated in a model with stochastic stirring and damping\cite{Takehiro:2007p577,Obuse:2010p578} that only forms jets.   As a technical necessity, both here and in the two-layer model analyzed later, an artificial viscosity called ``hyperviscosity,''  $\nu_2$, has been added to the equation of motion:
\begin{eqnarray}
\dot{q} = J[q, \psi] - \nu_2 \nabla^2 (\nabla^2 + 2) \zeta 
\label{barotropic-hyper}
\end{eqnarray}
Hyperviscosity differs from the usual physical viscosity term, $\nu \nabla^2 \zeta$, in having two more spatial derivatives.  Its presence signals a fundamental limitation of all numerical simulation of low viscosity fluids:  Fluctuations in the vorticity cascade down to smaller and smaller length scales, eventually dropping below the resolution of any numerical grid.   Hyperviscosity is a crude statistical model for the missing sub-grid scale physics, serving to absorb fluctuations only at the small length scales while at the same time hardly disturbing the total kinetic energy of the flow.  Hyperviscosity should also conserve angular momentum, and this accounts for the appearance of the $(\nabla^2 + 2)$ term that projects out the 3 modes with spherical wavenumber $\ell = 1$ that contain angular momentum.  This is a simple example of the {\it dynamical protection} of a conserved quantity. 

The {\it condensation} of vorticity into coherent structures (jets and vortices) occurs regardless of the detailed initial conditions, suggesting a simple variational interpretation.  The simplest interpretation is based upon a principle of minimum enstrophy (ME) as developed by Bretherton and Haidvogel\cite{Bretherton:1976p593} and by Leith\cite{Leith:1984p594}.  Enstrophy is the mean square vorticity, 
\begin{eqnarray}
\frac{1}{2}~ \int q^2~ d^2\Omega~ ,
\end{eqnarray}
and it (along with an infinite number of higher moments) is conserved in the absence of forcing or dissipation.   Hyperviscosity, however, drives enstrophy towards zero faster than the kinetic energy and this separation of time scales means that effectively an equilibrium state of well-defined energy can be reached.  This phenomenological approach captures in an intuitive fashion the physics of the inverse cascade process\cite{batchelor,Kraichnan:1980p227} whereby small-scale fluctuations organize into large-scale coherent structures.  

More rigorous statistical mechanical formulations have also been constructed to describe fluids in the limit of zero viscosity.  In such inviscid flows, finer and finer structures in the scalar vorticity field $\zeta(\vec{r})$ appear as the fluid evolves.  Miller and collaborators\cite{Miller90,Miller:1992p122} and Robert and Sommeria\cite{Robert91} (MRS) described the small-scale fluctuations of the fluid statistically using a local probability distribution $\rho(\vec{r}, \sigma)$ of vorticity  $\sigma$  over a small region of space centered at $\vec{r}$. The equilibrium state is then obtained by maximizing the Boltzmann entropy while conserving energy and the infinite number of other conserved quantities, namely the moments of the vorticity distribution.  In the physical limit of small but non-zero dissipation, MRS argued that the vorticity field is adequately captured by the coarse-grained mean field
\begin{eqnarray}
\bar\zeta(\vec{r}) = \int  \sigma \rho(\vec{r}, \sigma)~  d\sigma
\label{coarse}
\end{eqnarray}
of the corresponding zero-viscosity theory, with dissipation acting to smooth out the small-scale structures.  

However, that treatment of dissipation is unsatisfactory, because during the relaxation to equilibrium, viscosity can significantly alter the integrals of motion, especially the high-order moments.  Naso, Chavanis and Dubrulle\cite{naso:2010p467} showed that maximizing the MRS entropy holding only the energy, circulation, and enstrophy fixed is equivalent to minimizing the coarse-grained enstrophy as in the earlier ME theory.  The result uses the statistical MRS theory to justify the intuitive ME variational principle.  There is another problem with the statistical approach, however:  Figure \ref{figure3} shows that even at late times the state retains some memory of the initial condition.  For instance there is an isolated vortex in the southern hemisphere (somewhat like Jupiter with its Great Red Spot and jets\cite{cho1996}) and it remains there under further time integration.  The final state is determined by more than just the energy, the axial component of the angular momentum, and enstrophy.  

Weichman\cite{Weichman:2001,weichman05}, Majda and Wang\cite{majda06}, and Bouchet and his collaborators\cite{Venaille:2009p134,Bouchet:2010p364,Venaille:2010p513,Venaille:2010p514,Bouchet:2011p535}, have applied these ideas to single-layer models of geophysical flows such as Eq. \ref{barotropic}, finding a variety of coherent structures that switch between unidirectional jets and dipolar patterns, mimicking behavior observed in the seas.  (However Balk and co-workers have argued that an additional approximate invariant is an important ingredient in jet formation\cite{Balk:2011p580,Nazarenko:2009p133}.)  Verkley and Lynch have incorporated additional constraints and topography\cite{Verkley:2009p238}.  Near-equilibrium statistical descriptions of this type may be more applicable to oceans than to atmospheres because, as argued below, processes other than the inverse cascade are responsible for the formation of coherent structures in the atmosphere.

\section{Non-Equilibrium Physics By Direct Statistical Simulation}
\label{nonEquilibrium}

Inhomogeneous and anisotropic fluid flows pervade the universe.  From the jets and vortices in the atmospheres of the outer planets, to the differential rotation of the Sun, to the large-scale circulation of the Earth's atmosphere and oceans it is evident that such patterns of macroturbulence are the norm rather than the exception.  While anisotropy and inhomogeneity pose technical difficulties when it comes to mathematical descriptions, such flows appear to be amenable to DSS, because they provide the starting point for a perturbative expansion in fluctuations, as shown below.
DSS is able to reproduce statistics obtained by the traditional
route of time averaging DNS\cite{lorenz67}. DSS has
several advantages over DNS: (i) Low-order statistics are smoother in
space and stiffer in time than the underlying detailed flows.
Stationary fixed points or slowly varying statistics can therefore be described with
reduced degrees of freedom and can also be accessed rapidly. (ii)
Convergence with increasing resolution can be demonstrated, 
and for models with no subgrid physics, obviates
the need for separate closure models for unresolved processes.  Models with subgrid parameterizations are more naturally 
incorporated into DSS than DNS, because parameterizations are already inherently statistical.  
(iii) Finally and
most importantly, DSS leads more directly to insights, by integrating
out fast modes, leaving only the slow modes that contain the most
interesting information. This makes the approach ideal for simulating
and understanding decadal modes of the climate system, including
changes in these modes that are driven by climate change. The simplest
non-trivial DSS is based upon a second-order cumulant expansion that
can be improved systematically\cite{Marston:2008p1,Tobias:2011p550}.  Some related approaches can be found in Refs. \onlinecite{Canuto:1994p487,Canuto:2001p488}.  The Stochastic Structural Stability Theory (SSST) approach of Farrell and Ioannou\cite{Farrell:2007p611,Farrell:2009p237,Farrell:2009p636,Bernstein:2010p215} in particular is closely related to the cumulant expansion truncated at second order (CE2) discussed below.

One way to formulate the direct statistical approach is by Reynolds decomposing dynamical variables $q_i$ such as the vorticity into the sum of a mean value and a fluctuation (or eddy):
\begin{eqnarray}
q_i = \langle q_i \rangle + q_i^\prime\  \ {\rm with}~ \langle q_i^\prime \rangle = 0 
\label{ReynoldsDecomposition}
\end{eqnarray}
where the choice of the averaging operation, denoted by angular brackets $\langle \rangle$, depends on the symmetries of the problem.  Typical choices are time averages, or averages in the longitudinal direction (zonal averages), or averages over an ensemble of initial conditions. 
The first two equal-time cumulants $c_i$ and $c_{ij}$ of a dynamical field $q_i$ are defined by:
\begin{eqnarray}
c_i &\equiv& \langle q_i \rangle
\nonumber \\ 
c_{ij} &\equiv&  \langle q_i^\prime q_j^\prime \rangle = \langle q_i q_j \rangle - \langle q_i \rangle \langle q_j \rangle
\label{cumulants}
\end{eqnarray}
where the second cumulant contains information about {\it correlations} that are non-local in space, correlations that are called ``teleconnection patterns'' in the climate literature.

The models discussed here are members of a general class of quadratically non-linear systems that may be represented with coordinate-independent equation of motion:
\begin{eqnarray}
\dot{q}_i &=& A_i + B_{ij}~ q_j + C_{ijk}~ q_j q_k\ .
\label{algebraicEOMs}
\end{eqnarray}
Latin indices $i,~ j,~ k$ label coordinates that may represent discrete real-space points, modes in the space of spherical harmonics, or some other alternative basis.  The choice of an appropriate {\it basis} is of course of great practical importance in any actual calculation, just as for traditional condensed matter systems. There is an implicit sum over repeated indices.      

The equations of motion for the cumulants can be obtained directly by differentiating Eqs. \ref{cumulants} with respect to time and using Eqs. \ref{algebraicEOMs}, or {\it non-perturbatively} by introducing variables $p_i$ that are conjugate to the $q_i$ and defining the Hopf generating functional\cite{frisch95,ma:2005p108}:
\begin{eqnarray}
\Psi[q(t),~ p] &\equiv& e^{i p_i q_i(t)}\ .
\label{Psi}
\end{eqnarray}
The Hopf functional obeys a {\it many-body Schr\"odinger-like equation}:
\begin{eqnarray}
i \frac{\partial}{\partial t} \Psi &=& \hat{H} \Psi
\label{HopfEquation}
\end{eqnarray}
with linear operator $\hat{H}$ given by:
\begin{eqnarray}
\hat{H} &\equiv& p_i \left(-A_i + i B_{ij}~ \frac{\partial}{\partial p_j} + C_{ijk} \frac{\partial^2}{\partial p_j \partial p_k} \right)
\label{linearOp1}
\end{eqnarray}
as can be verified by combining Eqs. \ref{Psi}, \ref{HopfEquation}, and \ref{linearOp1} to reproduce Eq. \ref{algebraicEOMs}.
As Eq. \ref{HopfEquation} is linear in $\Psi$, the average $\overline{\Psi} \equiv \langle \Psi[q(t),~ p]  \rangle$ obeys the same equation; however $\overline{\Psi}$ now encapsulates information about the equal-time moments, as can be seen by repeated differentiation of Eq. \ref{Psi} with respect to $p_i$, followed by averaging:
\begin{eqnarray}
\langle q_{i_1} q_{i_2} \cdots q_{i_n} \rangle = (-i)^n~ \frac{\partial^n \overline{\Psi}}{\partial p_{i_1} \partial p_{i_2} \cdots 
\partial p_{i_n}} \bigg{|}_{p_i = 0}\ . 
\label{Hopf-moments}
\end{eqnarray}
The Hopf functional $\overline{\Psi}$ may also be expressed as the exponential of a power series in $p_i$, the coefficients being the cumulants:
\begin{eqnarray}
\overline{\Psi} = \exp \left\{ i c_i(t)~ p_i - \frac{1}{2!} c_{ij}(t)~ p_i p_j - \frac{i}{3!} c_{ijk}(t)~ p_i p_j p_k + \ldots \right\}
\label{Hopf-cumulants}
\end{eqnarray}
as can be checked by use of Eq. \ref{Hopf-moments} to reproduces the moments in terms of the cumulants, Eqs. \ref{cumulants}.  
Upon substituting Eq. \ref{Hopf-cumulants} into Eq. \ref{HopfEquation} and collecting powers of $p_i$ one obtains equations of motion for the cumulants that through third order read:
\begin{eqnarray}
\dot{c}_i &=& A_i + B_{ij}~ c_j + C_{ijk}~ (c_j~ c_k + c_{jk})
\nonumber \\
\dot{c}_{ij} &=& \{2 B_{ik}~ c_{kj} + C_{ik\ell}~ (4 c_\ell~ c_{jk} + 2 c_{jk\ell})\}
\nonumber \\
\dot{c}_{ijk} &=& \{3 B_{i \ell}~ c_{\ell j k} + 6 C_{k \ell m}~ (c_{ijm}~ c_\ell + c_{im}~ c_{j \ell})\} - \mu~  c_{ijk} + {\cal O}(c_{ijk\ell})\ .
\label{ce-eom}
\end{eqnarray}
Here for compactness the short-hand notation $\{ \}$ has been used to denote symmetrization over indices.  For example:
\begin{eqnarray}
\{2 B_{ik}~ c_{kj}\} \equiv  B_{ik}~ c_{kj} + B_{jk}~ c_{ki}
\end{eqnarray}
maintains the symmetry $\dot{c}_{ij} = \dot{c}_{ji}$.  

\begin{figure}
\centerline{\includegraphics[width=5in]{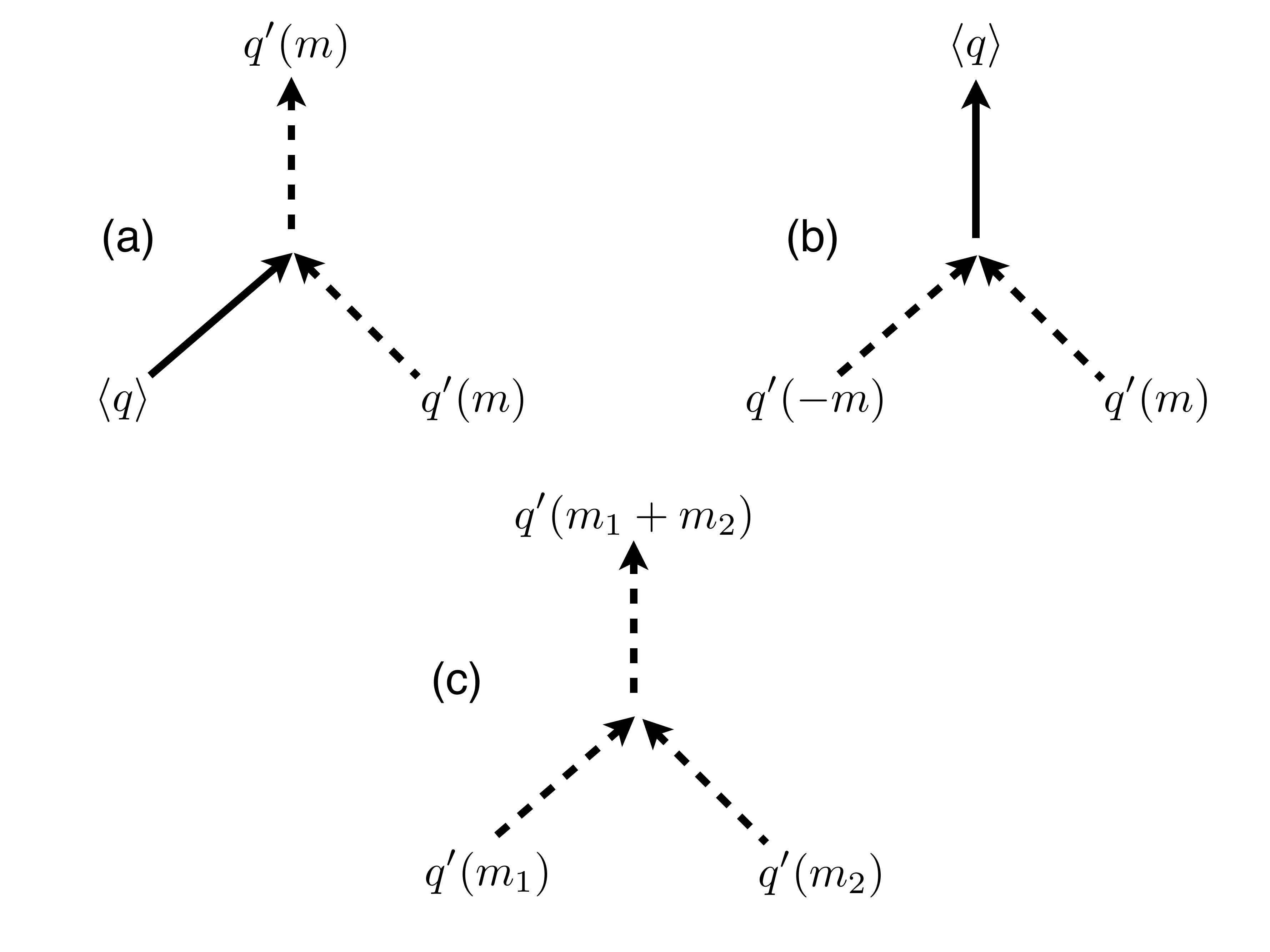}}
\caption{\label{figure4} Eddy-mean flow and eddy-eddy triad interactions. (a) An eddy of zonal wavenumber $m \neq 0$ scatters off the mean flow.  (b) Two eddies of equal but opposite zonal wavenumber combine to alter the mean flow.  This is the Reynolds forcing.  (c) The eddy-eddy interaction, neglected at level CE2, is included at levels CE3 and higher.}
\end{figure}

Processes involving the non-linear term $C_{ijk}$ in Equations \ref{ce-eom} can be visualized diagrammatically.  The term $C_{ijk}~ c_{jk}$ in the first of the equations represents the contributions of eddies to the mean flow known as Reynolds forcing (Figure \ref{figure4} (b)).  Scattering of an eddy by the mean flow is captured by the term $4~ C_{ik\ell}~ c_\ell~ c_{jk}$ in the second equation and is depicted by Figure \ref{figure4} (a).  Finally, eddy-eddy scattering occurs only at CE3 and higher levels (the term $6~ C_{k \ell m}~  c_{im}~ c_{j \ell}$ in the last of Equations \ref{ce-eom} that is represented by Figure \ref{figure4} (c)).  

Truncated at second-order (CE2) the cumulant expansion is realizable\cite{salmon98} (CE2 can be viewed as the exact solution of a stochastically-driven linear model) and well-behaved in the sense that the energy density is positive and the second cumulant obeys positivity constraints.  The closure amounts to the neglect of eddy-eddy
interactions\cite{Bernstein:2010p215,Tobias:2011p550}.  CE2 retains the eddy-mean flow interaction. Two eddies of equal but opposite zonal wavevectors can also 
combine to alter the mean-flow by the Reynolds mechanism.  
Mathematically CE2 amounts to the assumption that the probability 
distribution function is adequately described by a normal, or Gaussian, distribution.
Cumulant expansions are known to fail badly for the difficult
problem of homogeneous, isotropic, 3D turbulence\cite{frisch95}. In
the case of anisotropic and inhomogeneous geophysical flows, however,
even approximations at the CE2 level can do a surprisingly
good job of describing the large-scale circulation qualitatively.  Closing the equations of motion at third-order (CE3) and beyond introduces theoretical difficulties.  For instance a phenomenological eddy-damping parameter\cite{Orszag:1977,Andre:1974p398} $\mu$ that models the neglect of the fourth and higher cumulants from the hierarchy has been included in the last of Eq. \ref{ce-eom} and is needed to prevent the time-evolution of the cumulants from diverging.  In practice, however, the statistics are insensitive to the value of $\mu$ and as shown below CE3 can improve the quantitative agreement with DNS, and even repair qualitative problems with CE2.  In the absence of forcing and dissipation, both CE2 and CE3 conserve the same quantities as DNS: angular momentum, mean potential temperature, total energy, and mean-square potential temperature.  Thus systematic expansions in the retained cumulants yield {\it conserving approximations}, a highly desirable trait as these help to protect against pathologies introduced by the truncation. 

Still unsolved is the problem of how to systematically treat non-linearities that are not low-order polynomials (Eq. \ref{algebraicEOMs}), such as the non-linearity due to latent heat release that is described by a step-function.  Such non-linearities can be incorporated at the CE2 level by using the fact that the probability distribution function is a Gaussian\cite{OGorman:2006p436} but it is not clear how to go beyond CE2.  As nonlinearities more complicated than quadratic ones appear in parameterizations of boundary layers, convection, and radiation, this is an important open question that needs addressing.

\subsection{Relaxation to a Prescribed Jet}
\label{Jet}

\begin{figure}
\centerline{\includegraphics[width=6in]{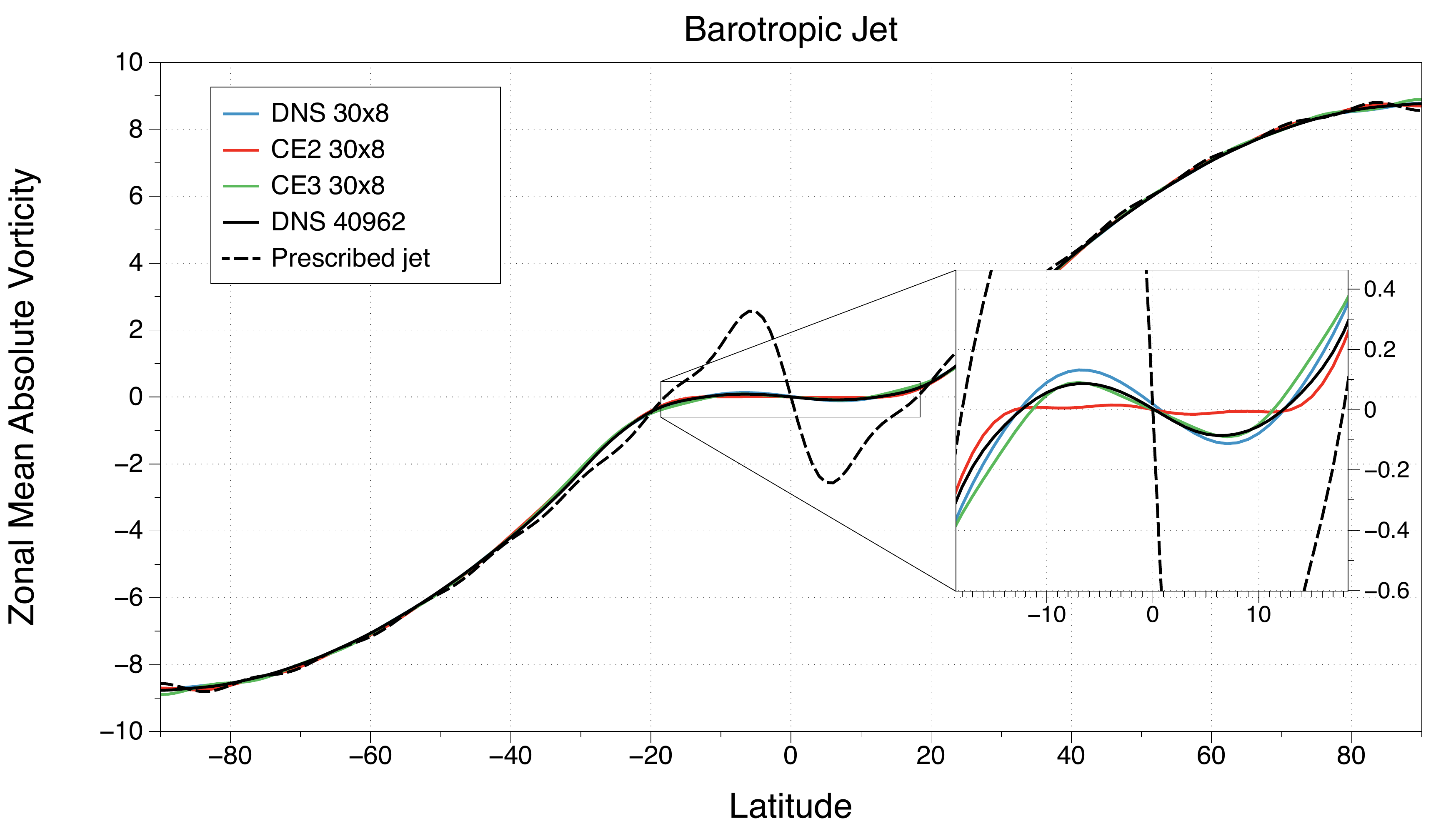}}
\caption{\label{figure5} Zonal mean absolute vorticity of the barotropic jet as accumulated by DNS in spectral space (truncated to modes with $\ell \leq 30$ and $m \leq 8$) and compared to DSS by second- and third-order cumulant expansions (CE2 and CE3) with the same spectral resolution.  Turbulence in the mixing region around the equator flattens the mean vorticity profile away from that of the prescribed jet $q_{\rm jet}$ (dashed line).  The rotation period is one day ($\Omega = 2 \pi/$ day), $\Gamma = 0.6 \Omega$, $\Delta \phi = 0.05$ and $\tau = 50$ days (see Equations \ref{barotropicJet} and \ref{prescribedJet}).    Also shown for comparison are statistics accumulated from DNS on a spherical geodesic grid with $40,962$ cells, demonstrating that the spectral truncation employed here is not a severe approximation.}
\end{figure}

\begin{figure}
\centerline{\includegraphics[width=5in]{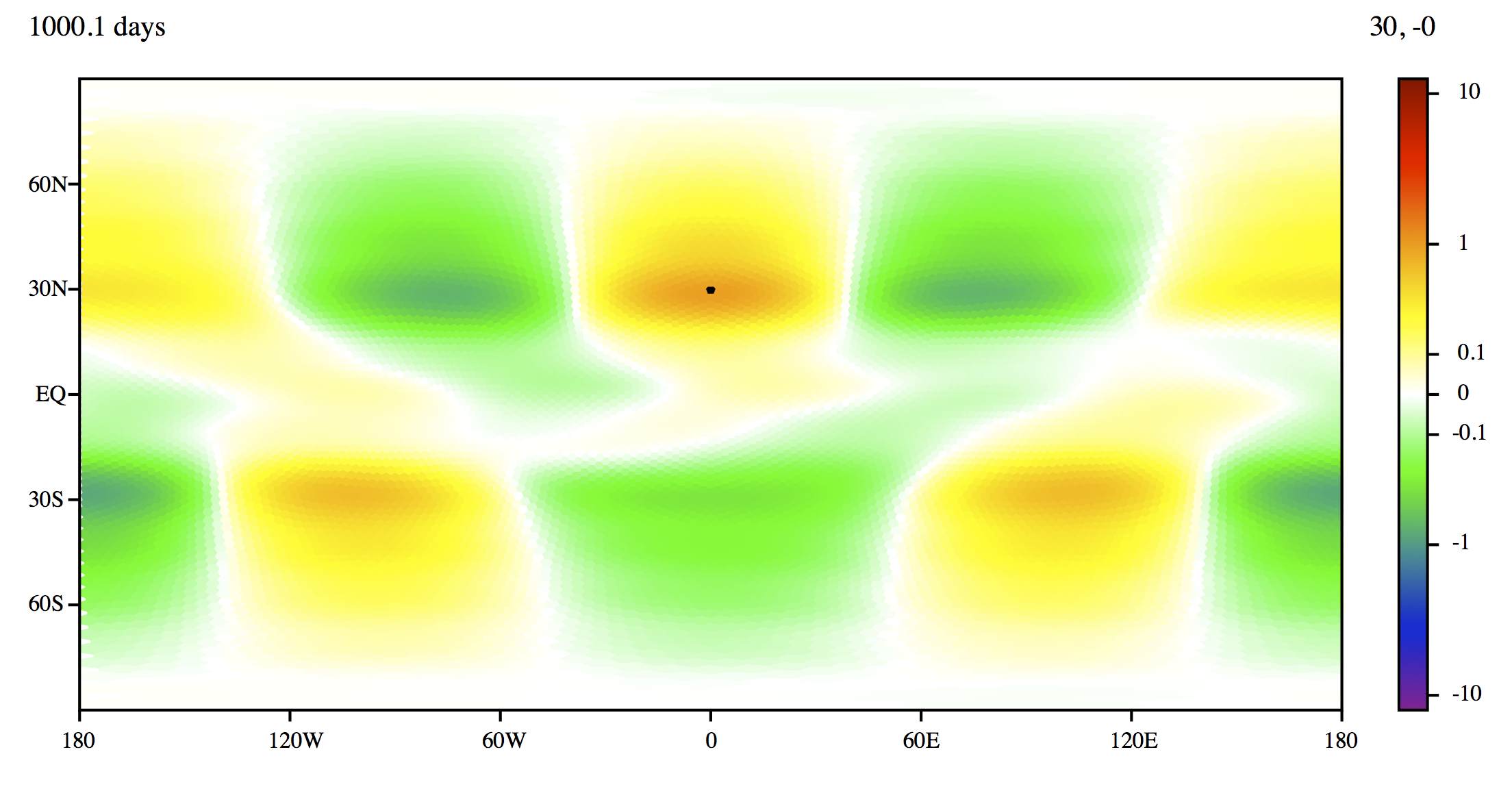}}
\vskip 0.5cm
\centerline{\includegraphics[width=5in]{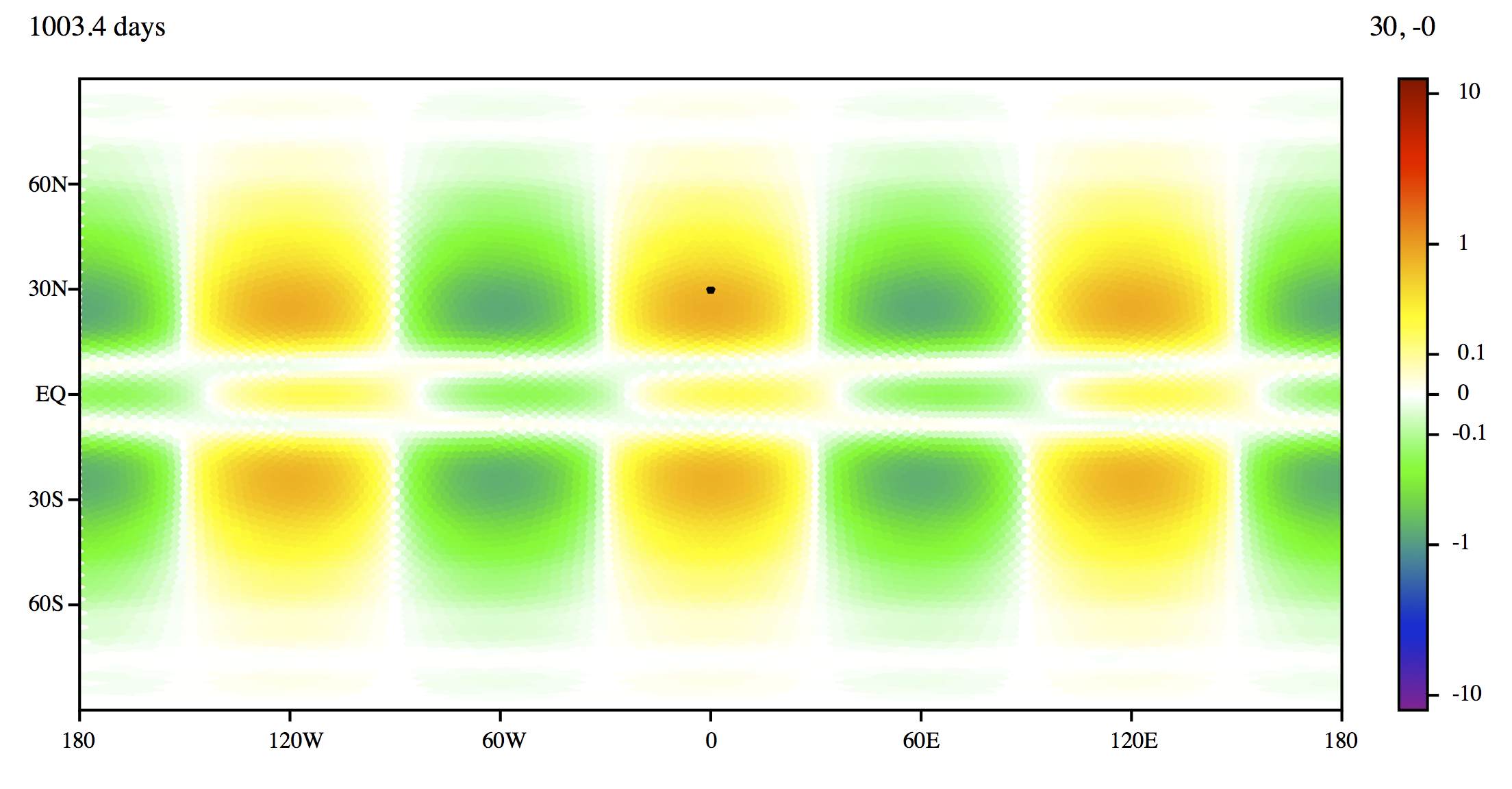}}
\vskip 0.5cm
\centerline{\includegraphics[width=5in]{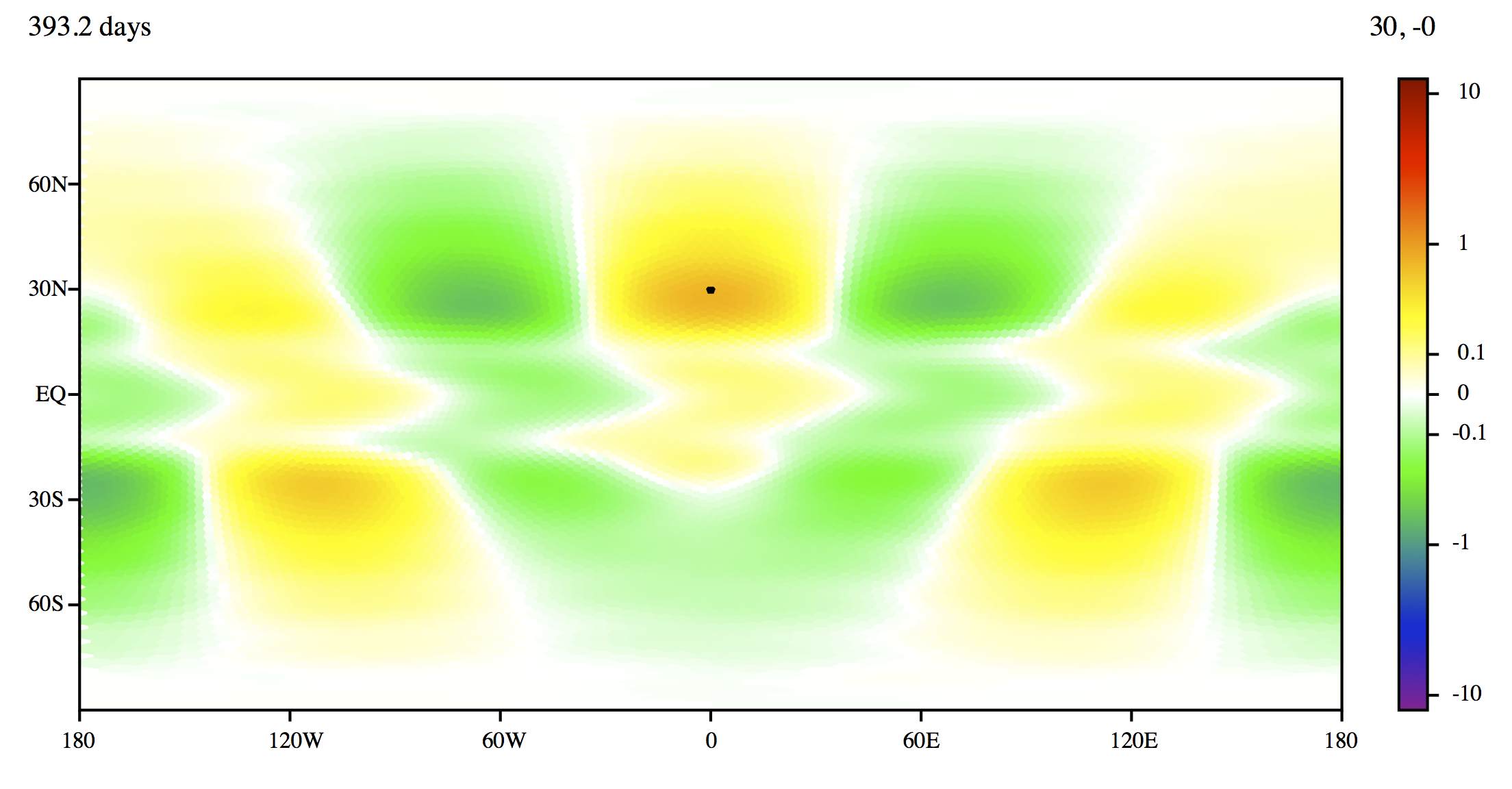}}
\caption{\label{figure6} Second cumulant of vorticity as calculated by DNS (top), CE2 (middle) and CE3 (bottom).  The two-point correlations are with respect to a reference point at a latitude of $30^\circ$ and along the prime meridian marked by the black dot.  The correlations show that the dominant fluctuations have zonal wavenumbers $2$ and $3$.  CE2, however, only exhibits fluctuations at wavenumber $3$.  This defect is repaired at the CE3 level.}
\end{figure}

The efficaciousness of DSS by expansions in cumulants can be illustrated by a simple single-layer model of barotropic flow coupled to an underlying prescribed point jet, a classic prototypical system\cite{lindzen83,Nielsen84,schoeberl84,shepherd:1988p548}.  The equation of motion is:
\begin{eqnarray}
{{\partial q}\over{\partial t}} + J[\psi, q] = \frac{q_{\rm jet} - q}{\tau}\ .
\label{barotropicJet}
\end{eqnarray}
Forcing and dissipation are represented by the term on the right-hand-side of
Eq.~(\ref{barotropicJet}), which linearly relaxes the absolute vorticity $q$ to
that of a prescribed zonal jet $q_{\rm jet}$ over a relaxation
time a $\tau$.  The prescribed jet is symmetric about the equator, with retrograde (east to west) velocity
peaking at the equator.  Such flow is described by 
constant relative vorticities $\pm\Gamma$ on the flanks far away from
the apex and by a rounding width $\Delta \phi$ of the apex,
\begin{eqnarray}\label{q_jet}
q_{\rm jet}(\phi) = f(\phi) - \Gamma \tanh\left(\frac{\phi}{\Delta \phi}\right)\ 
\label{prescribedJet}
\end{eqnarray}
and appears as the dashed line in Fig. \ref{figure5}.
Flow coupled to the jet becomes unstable and turbulent at sufficiently large values of $\tau$.  
As shown in Fig. \ref{figure5} the absolute vorticity mixes in the equatorial region, flattening out the profile, an effect that is qualitatively captured by CE2\cite{Marston:2008p1}.  However CE2 overestimates the degree of mixing; by contrast CE3 matches statistics accumulated by DNS rather well quantitatively.  Examination of the second cumulant two-point correlation function (Figure \ref{figure6}) shows that while CE2 now fails even qualitatively to describe the statistics as determined by DNS, CE3 does rather well.  The eddy-eddy interaction retained within CE3 mixes modes with different zonal wavenumbers.  Wavenumbers $2$ and $3$ dominate, yielding a pattern similar to the one seen in DNS.  
The eddy diffusivity finds is negative in some regions of the jet, both in DNS and in the cumulant expansions\cite{Marston:2008p1}.  This demonstrates that naive (but common) closures that replace non-linear advection with local linear diffusion must break down.  

Stratification and shearing act together to weaken the nonlinearities in large-scale flows by pulling apart vorticity before it can accumulate into highly nonlinear eddies.  At the same time the eddies act {\it collectively} back upon the mean-flow by the Reynolds forcing mechanism, altering it\cite{Tobias:2011p550}.  It is these features of the general circulation that distinguish it from the highly nonlinear problem of 3D isotropic and homogenous turbulence\cite{Schneider2006} and make viable a systematic perturbative expansion in fluctuations about the mean-flow.  

\subsection{Two Layer Primitive Equation Model}

\begin{figure}
\centerline{\includegraphics[width=5in]{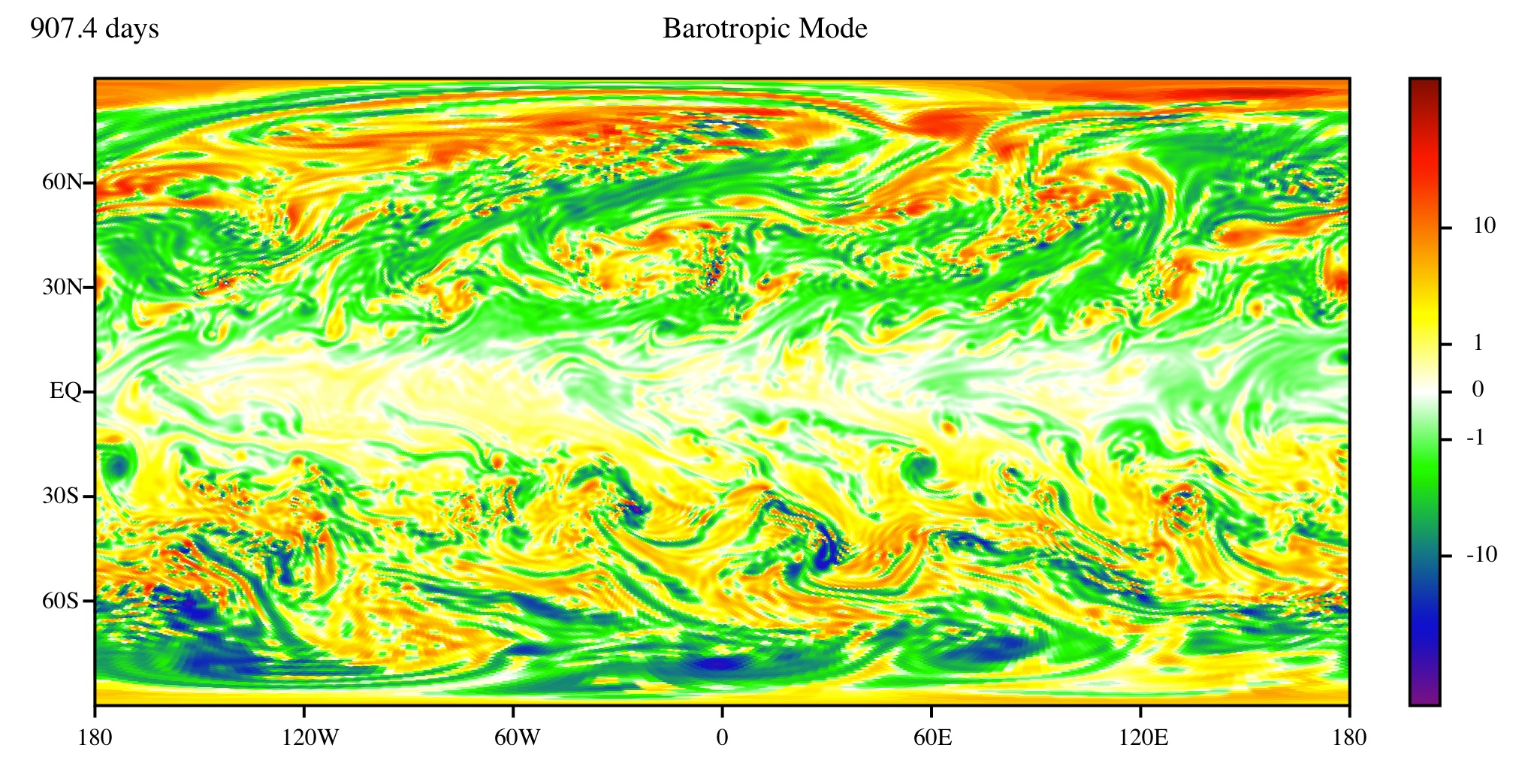}}
\vskip 0.2cm
\centerline{\includegraphics[width=5in]{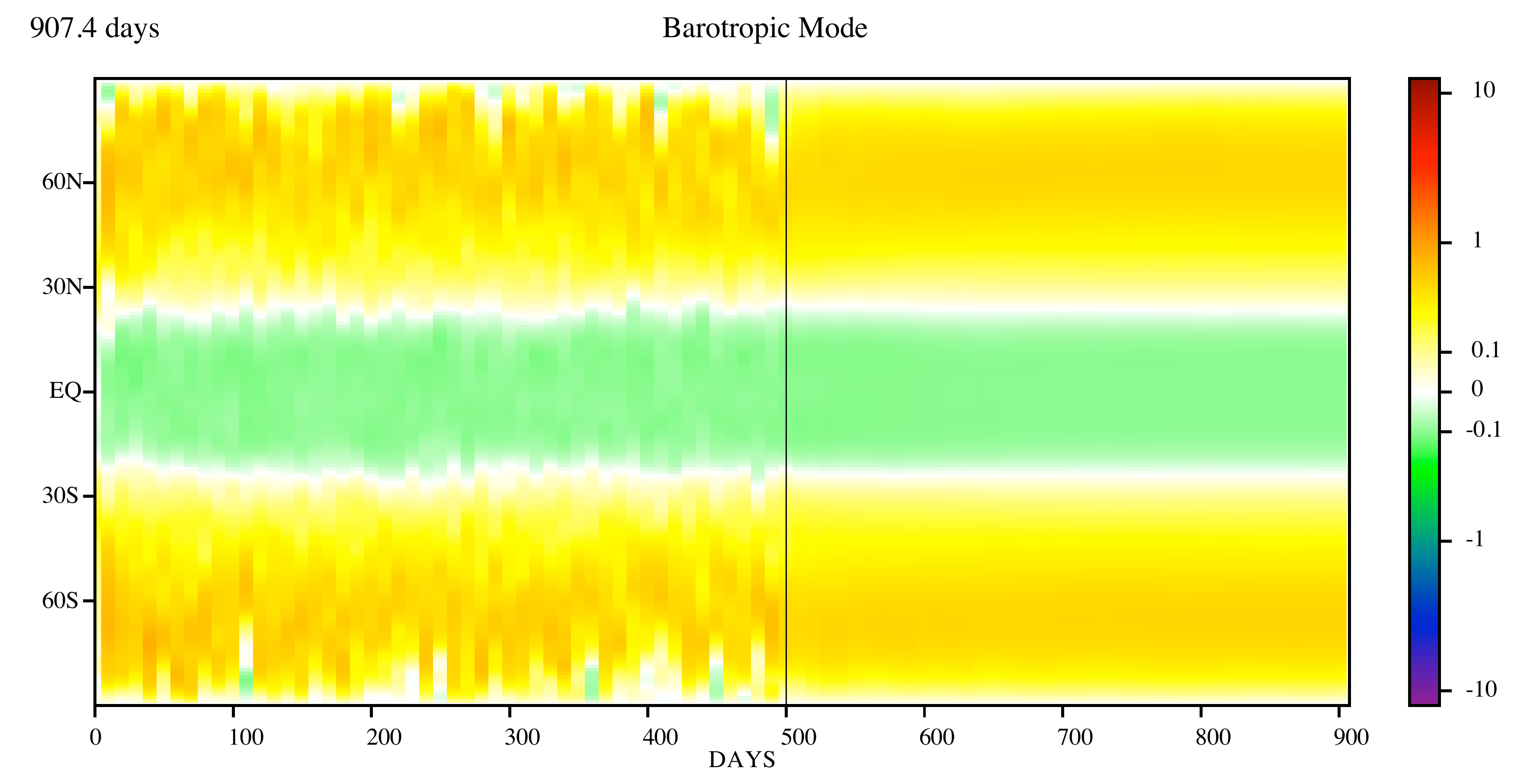}}
\vskip 0.2cm
\centerline{\includegraphics[width=5in]{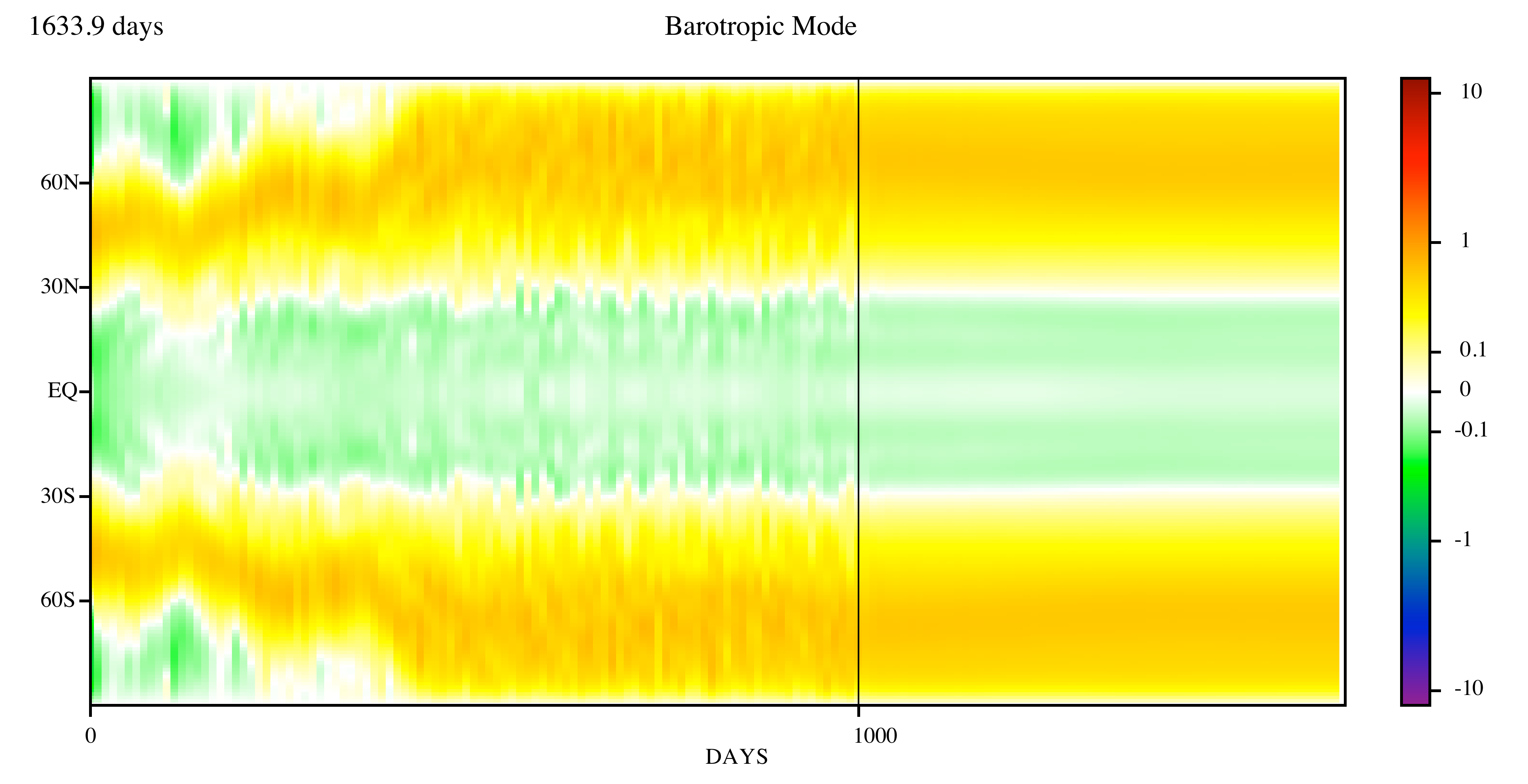}}
\caption{\label{figure7} Top: Snapshot of the barotropic vorticity as calculated by DNS on a spherical geodesic grid of $163,842$ cells.  Here the rotation period is one day, $a = 1$ Earth radius, $\tau_R = 33$ days, $1/\kappa = 10$ days, the barotropic potential temperature relaxes to a $\overline{\theta}_0 = 120$K equator-to-pole gradient, and the baroclinic temperature relaxes to $\hat{\theta}_0 = 30$K, suppressing convection.  Middle and bottom: Timelines of the mean zonal velocity, averaged over both layers, as calculated by DNS (middle panel) and by CE2 with spectral truncation $\ell \leq 50$ (bottom panel).  Time averaging is turned on at the time indicated by the black vertical line, smoothing out any fluctuations that survive the zonal average.  The mid-latitude westerly winds and the tropical easterlies are similar to Earth's general circulation.}
\end{figure}

\begin{figure}
\centerline{\includegraphics[width=6in]{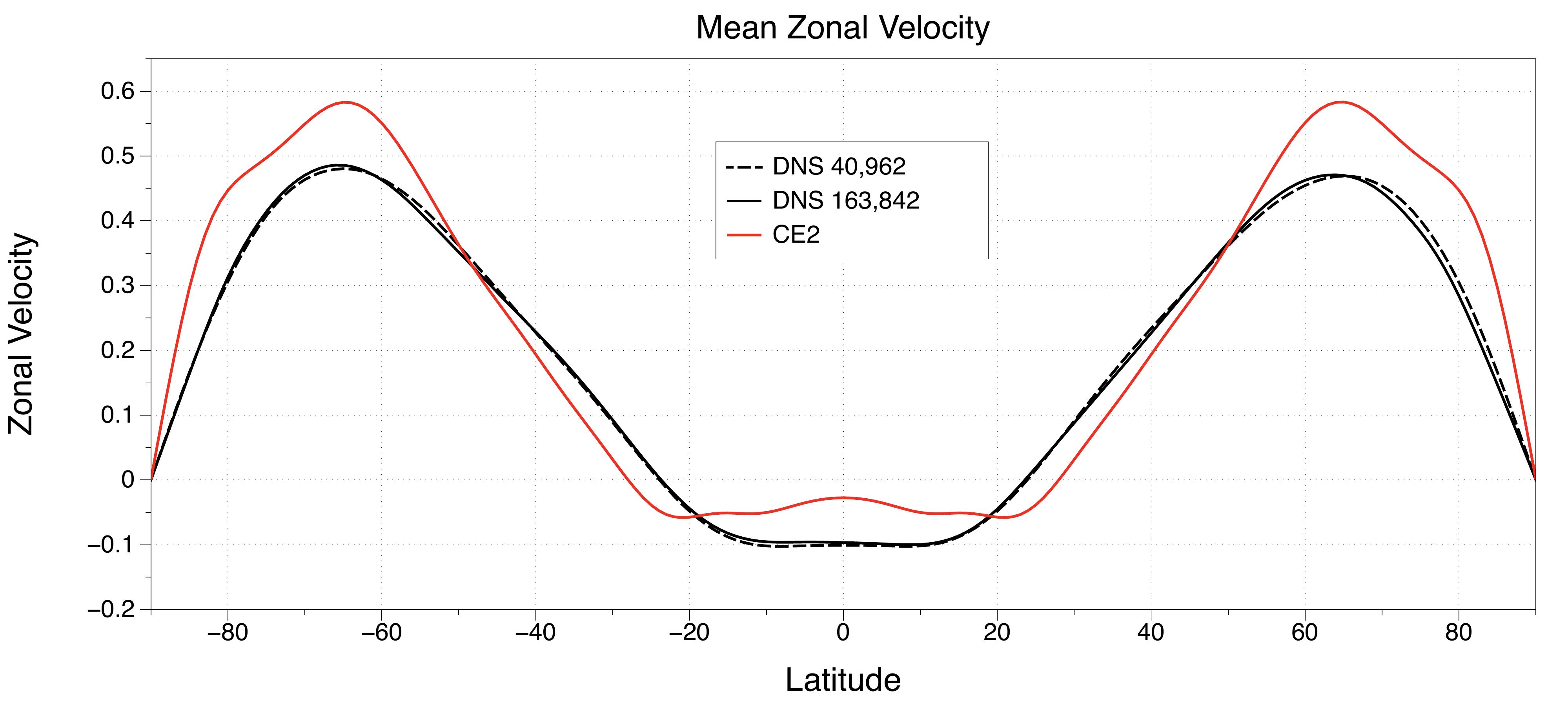}}
\caption{\label{figure8} Mean zonal velocity, averaged over both layers, for the two-layer primitive equation as accumulated by DNS on spherical geodesic grids of two different resolutions and compared to CE2.  Convergence in the statistic with increasing DNS resolution is apparent.  Both DNS and CE2 reproduce midlatitude westerlies and tropical easterly winds, but CE2 only agrees with DNS at a qualitative level.  Same parameters as Figure \ref{figure7}.}
\end{figure}

A qualitatively realistic model of the general circulation, at least for regions outside of the tropics that are dominated by convection, is provided by the two layer primitive equation model introduced in subsection \ref{two-layers}.  As shown in Figure \ref{figure7} the model develops midlatitude westerly jets and storm tracks, as well as easterly trade winds in the tropics.  Hadley cells with overturning circulation near $\pm 30^\circ$ latitude mimic those of Earth.  Zonal means determined by CE2 agree qualitatively with DNS, but as Figure \ref{figure8} shows, there are quantitative discrepancies in the zonal winds, especially at high latitudes.  

\begin{figure}
\centerline{\includegraphics[width=5in]{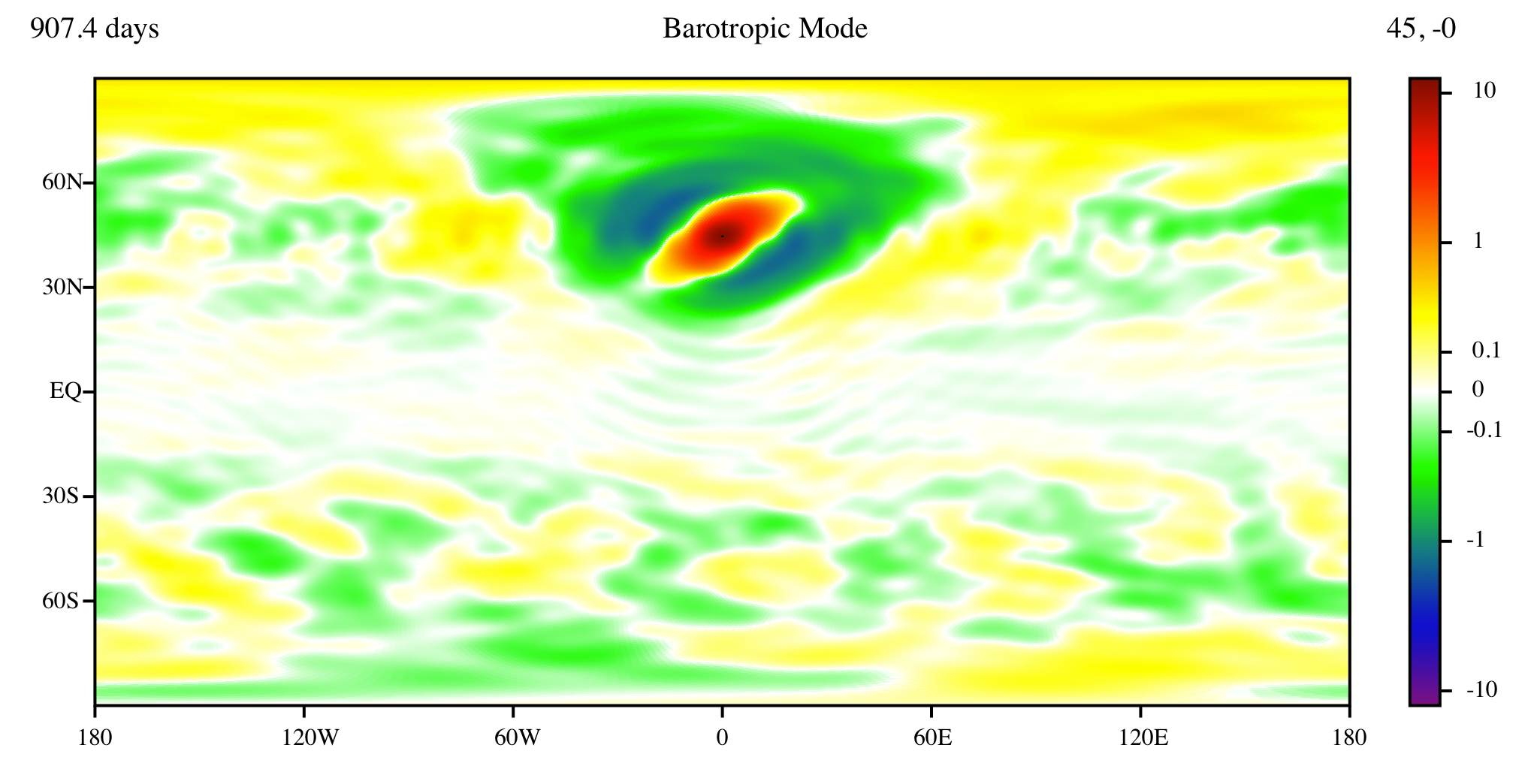}}
\vskip 0.5cm
\centerline{\includegraphics[width=5in]{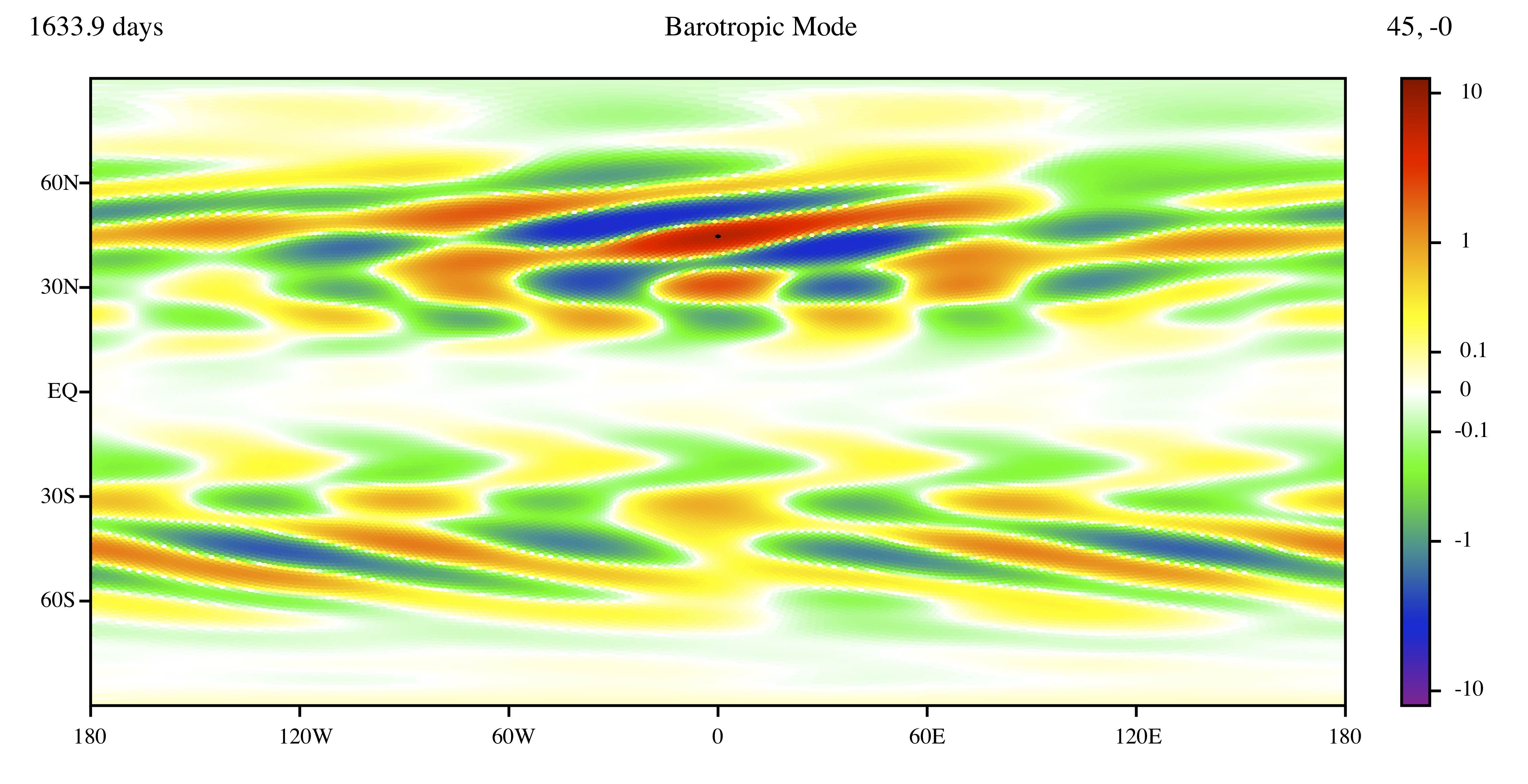}}
\caption{\label{figure9} Second cumulant of barotropic vorticity as calculated by DNS on a spherical geodesic grid of $163,842$ cells (top) and CE2 retaining modes with $\ell \leq 50$ (bottom).  The two-point correlations are with respect to a reference point near the middle of the storm track at a latitude of $45^\circ$ and along the prime meridian. Same parameters as Figure \ref{figure7}.  Correlations in the flow are seen to be much shorter ranged in DNS than in CE2.}
\end{figure}

The second cumulant two-point correlation function is compared in Figure \ref{figure9}.  Correlations are considerably shorter-ranged in DNS than in CE2, and the correlations are confined to a single hemisphere.  Whether or not CE3 would address these discrepancies is an open question.   
In any case, longer ranged correlations are evident both in a two-layer quasi-geostrophic model\cite{Marston:2010} and in a multilayer model\cite{OGorman:2007p76} so the short-ranged nature of the correlations evident here in DNS may be an artifact of the two-layer approximation.  

Additional insights into the general circulation, and the formation of zonal jets, can be gained from the spectrum of kinetic energy fluctuations.  In both simulations of the general circulation with and without the eddy-eddy interaction\cite{OGorman:2007p76}, and in the real atmosphere\cite{Boer1983,Nastrom1985}, there is a $\ell^{-3}$ power-law {\it scaling} for spherical wavenumbers $\ell$ greater than the characteristic eddy wavenumber of approximately $10$.  Frequently this power-law decay is explained by invoking a cascade of enstrophy towards small scales\cite{Kraichnan:1980p227,salmon98}; however no such cascade can occur in the absence of eddy-eddy interactions.  Furthermore, in the real atmosphere there is no inertial range for which eddy-eddy interactions dominate\cite{OGorman:2007p76}.  Rather other contributions to the spectrum stemming from the conversion of potential to kinetic energy, and friction, are important\cite{Lambert1987,Straus1999}, making arguments for power-law scaling that are based upon enstrophy cascades questionable\cite{Vallis:1992}.  
 
\subsection{Velocity Profile of a Vortex}
In the absence of the Coriolis force, large-scale vortices condense in two-dimensional flows\cite{batchelor,Kraichnan:1980p227,cho1996,Chertkov:2007p589,Marston:2011p604}.  Chertkov, Kolokolov, and Lebedev\cite{Chertkov:2010p588} have used an expansion similar to CE3 to study the mean radial profile of the velocity of an isolated vortex.  They were able to reproduce the $r^{-\frac{1}{4}}$ power-law decrease in the velocity as a function of the distance $r$ from the vortex center seen in DNS\cite{Chertkov:2007p589}.  Here, as in the previous examples, the strong mean-flow around the vortex provides a basis for a perturbative expansion in fluctuations about it.  
 
\section{Other Approaches}
\label{otherApproaches}

The complexity of non-equilibrium statistical mechanics raises the question of whether or not simple principles can be found that generalize equilibrium statistical mechanics and thermodynamics to non-equilibrium systems.  Two such approaches are outlined below.  

\subsection{Maximum Entropy Production}
\label{MEP}

\begin{figure}
\centerline{\includegraphics[width=6in]{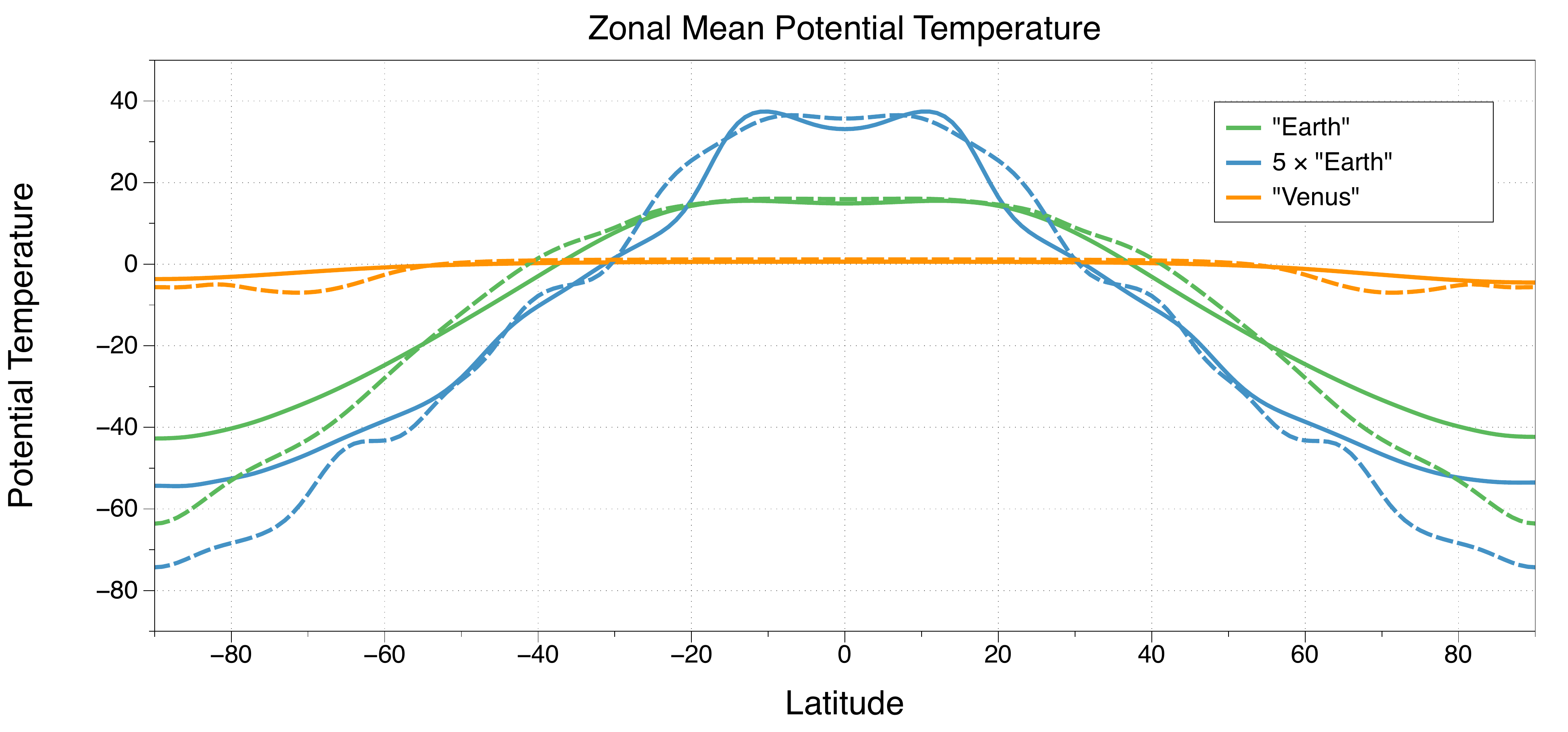}}
\vskip 0.5cm
\centerline{\includegraphics[height=3.5in]{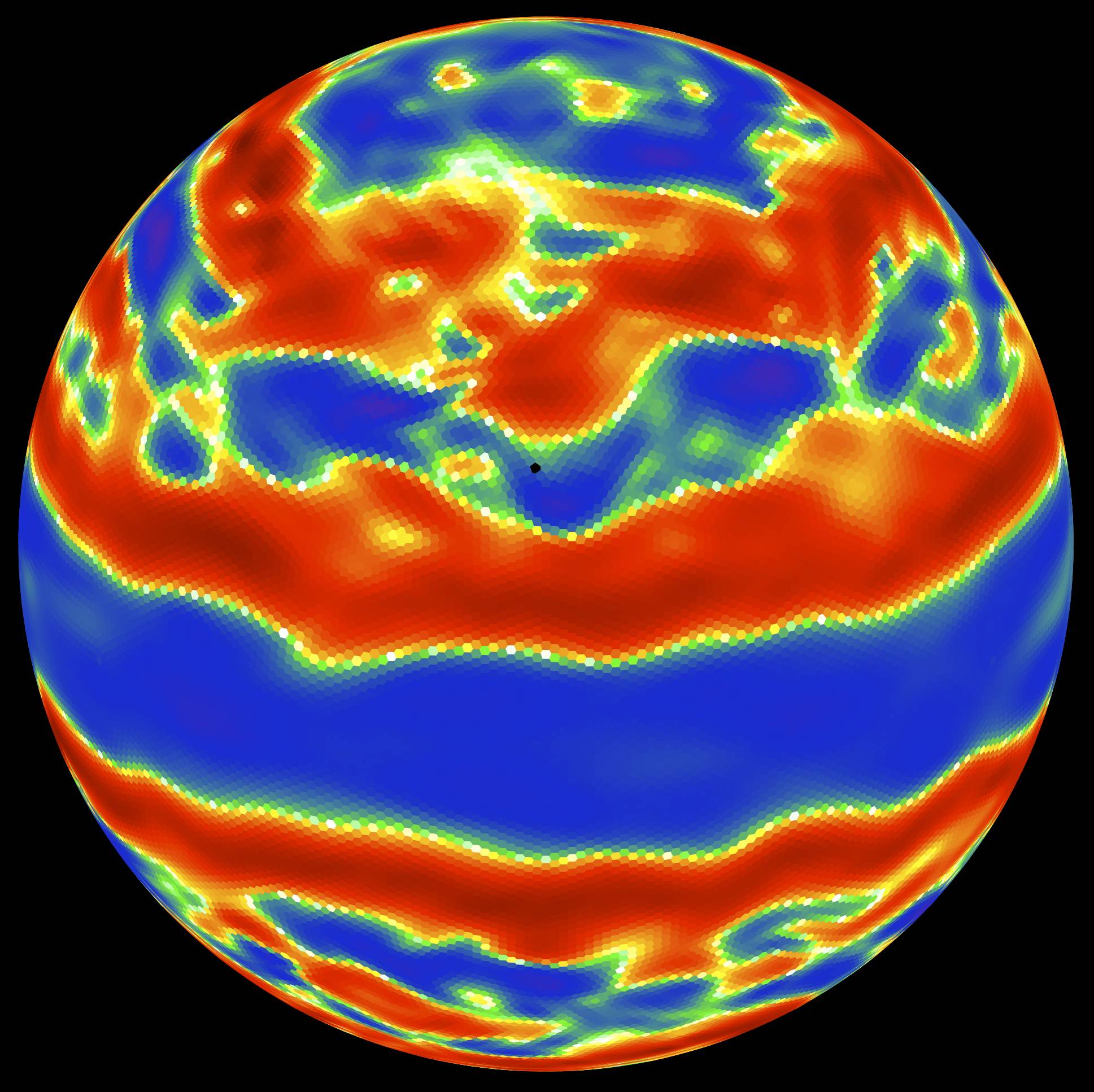} \includegraphics[height=3.5in]{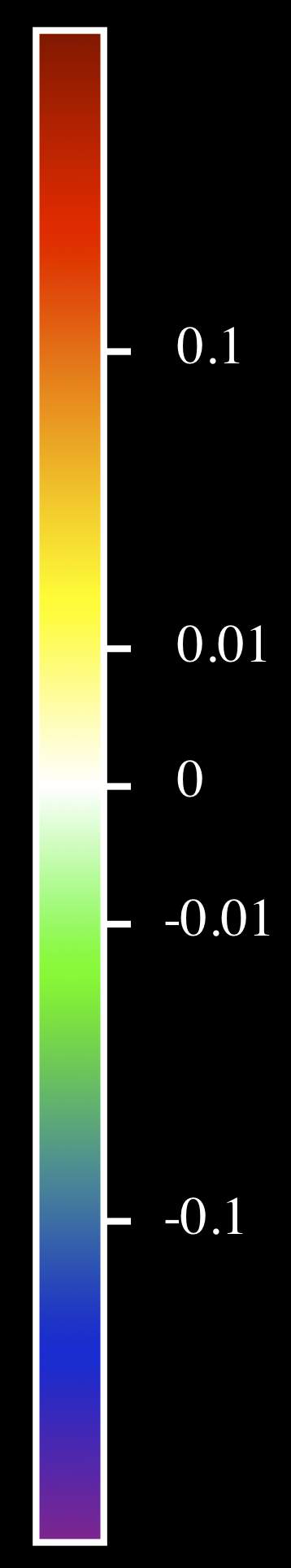}}
\caption{\label{figure10}Top: Time and zonally averaged potential temperature as calculated by DNS (solid lines) and CE2 (dashed lines) of the two-layer primitive equation model for three planets that differ only in their rate of rotation.  In each case the atmosphere relaxes to an imposed $120$ K equator-to-pole temperature difference.  ``Earth'' rotates once per day; ``Venus'' has a rotation period of 100 days, and ``5 $\times$ Earth'' completes a rotation in 0.2 days.  Bottom: A snapshot of the zonal velocity for ``5 $\times$ Earth'' reveals that it has two westerly jets in each hemisphere.  Except for the rotation rate, the parameters are the same as Figure \ref{figure7}.}
\end{figure}

A school of thought has built up around the provocative notion that a principle of maximum entropy production (MEP) offers a route to understanding systems driven away from equilibrium comparable to the Boltzmann hypothesis in equilibrium statistical mechanics.  Refs. \onlinecite{Lorenz:2003p210,Ozawa:2003p212,Kleidon:2004p208,Whitfield:2005p166,Woollings:2006p162,Goody:2007p163,Lucarini:2009p688} are a sample of recent work. An early apparent success came when Paltridge applied these ideas to simple box models of the Earth's atmosphere.  Paltridge's model assumed that the atmosphere acts either to maximize the production of entropy, or alternatively maximize dissipation, as it {\it transports} heat poleward. Intriguingly the calculated pole-to-equator temperature difference was close to the observed difference of about 60 K\cite{Paltridge:1975,Paltridge:1979p470,Paltridge:2001}.   However the planetary rotation rate does not enter the calculation -- yet it certainly exerts strong control over the temperature gradient.  On Earth the Coriolis force strongly deflects winds in meridional directions, inhibiting the transfer of heat out of the tropics.  On a planet rotating more rapidly than Earth there should be an even larger equator-to-pole temperature gradient; conversely on a slowly rotating planet the atmosphere would efficiently redistribute heat, nearly erasing the gradient.  The two-layer primitive equation model illustrates the strong dependence on the rotation rate nicely (see Figure \ref{figure10}).  That Paltridge's idea yielded a pole-to-equator temperature difference close to reality appears to be a coincidence.   

Dewar has attempted to ground MEP in more fundamental principles\cite{Dewar:2003p207,Lorenz:2003p210,Dewar:2005p211}.    Dewar used the maximum information entropy approach of Jaynes (see his two 1957 papers published in {\it Physical Review} \cite{Jaynes:1957a,Jaynes:1957b}) to generalize the equilibrium ensemble by introducing a probability measure over space-time paths taken by systems that are driven away from equilibrium.  However a number of problems with Dewar's theory have been discussed in the literature\cite{Grinstein:2007p419,Bruers:2007p457,Paquette:2010p520}.  
In particular Paquette\cite{Paquette:2010p520} considered an exactly solvable Langevin equation that describes a massive Brownian particle in a viscous medium, driven by stochastic fluctuations resulting from the impact of many light particles.  He showed that the probability functional for a given particle path cannot be reproduced by application of Dewar's framework\cite{Dewar:2003p207}.  As stochastic models are a common way to drive large-scale motion in reduced models\cite{Chandrasekhar:1943p674,vandenEijnden:1998p676}, and also find increasing use in weather and climate simulations as a way to model unresolved processes such as small-scale convection\cite{Hasselmann:1976p675,Palmer:2010}, the failure of the Dewar formalism to yield the correct statistics raises the question of its wider applicability.   

The two-layer model also reveals other interesting phenomena such as the {\it spontaneous breaking} of time-reversal symmetry.  For a planet with no rotation, $\Omega = 0$ and simulation finds no preference for the atmosphere to rotate in a particular sense.  The total angular momentum of the atmosphere remains zero.  A small but non-zero planetary rotation rate is enough to induce a striking superrotation of the atmosphere, with vigorous winds at all latitudes moving in the direction of planetary rotation, but moving much faster than the planet's surface.  The Hadley cells now extend over each entire hemisphere.  Such superrotation has been seen in the winds of Venus and Titan.  

In the opposite limit of high rotation (the $5 \times$ Earth model in Figure \ref{figure10}) the system undergoes a {\it phase transition} with multiple westerly jets now appearing in each hemisphere.  Multiple jets are also found in the atmosphere of Jupiter and can be understood in terms of the Rhines scale
\begin{eqnarray}
\ell_R &\equiv& \sqrt{u_{rms} / \beta}
\end{eqnarray}
where
\begin{eqnarray}
\beta &\equiv& \frac{d f}{d (a \phi)}
\end{eqnarray}
is the gradient of the Coriolis parameter, $u_{rms}$ is the root mean square speed of the jet eddies\cite{tapio}, and $a$ is the radius of the planet.
The Rhines scale determines the width of the jets.  Increasing $\Omega$ increases $\beta$ and makes $\ell_R \ll a$, permitting more than one jet to fit into each hemisphere\cite{Galperin:2007p595}.  SSST is able to reproduce the linear dependence of the number of jets on the inverse Rhines scale\cite{Farrell:2007p611}.

\subsection{Linear Response}
\label{linearResponse}
The fluctuation-dissipation theorem relates the {\it linear response} of a system to perturbing influences to two-point correlations (the fluctuations).  If generalizable to non-equilibrium systems, an obvious application would be to understand the influence of various drivers of climate change, such as rising levels of greenhouse gases, upon the climate within such a framework.  The approach has been applied to highly simplified models with just a few degrees of freedom\cite{Cooper:2011p640,KirkDavidoff:2009p629}.  One lesson learned from the work of Kirk-Davidoff\cite{KirkDavidoff:2009p629} is that care must be taken when applying linear response theory to actual climate data\cite{Schwartz:2007p616} to extract an estimate of climate sensitivity, as long time series are necessary\cite{Scafetta:2008p617}.   Application of linear response theory to a full atmospheric general circulation model with many degrees of freedom has also been attempted by 
Gritsun, Branstator, and Majda\cite{Gritsun:2008p277}.  These authors report qualitative agreement between the linear and the actual response of the model.

\section{Summary and Open Questions}
\label{conclusion}

Viewing planetary atmospheres from the perspective of condensed matter physics provides new insights.   Idealized models of atmospheres, like those of traditional condensed matter, can be understood in detail and used to explain observed phenomena such as the statistics of the general circulation.  Ideas and tools discussed here may see application to both deep-time reconstructions of climate (as ESMs have difficulty simulating vast periods of time) and to newly discovered and characterized exoplanets, as well as to the present-day climate of Earth and the pressing question of climate change.  

Still unclear, however, is the range of applicability of such approaches, and whether or not they can advance from being qualitative tools to become quantitatively accurate theories of the general circulation, possibly even replacing the traditional technique of direct numerical simulation.  There is thus a pressing need for improved theoretical frameworks and tools to better describe non-equilibrium statistical physics.    Will near-equilibrium theories prove valuable in the description of real geophysical flows?  Are there reliable general principles of non-equilibrium statistical physics waiting to be discovered?  Can quantitatively accurate DSS closures be found that are at the same time computationally efficient?  Can non-perturbative formulations be implemented, possibly with the use of renormalization-group ideas?\cite{mccomb2004}  Is it possible to extend DSS to ESMs that are complex enough to address questions such as how the storm tracks and ocean currents will shift as the climate changes?\cite{OGorman:2010p506}     By thinking about atmospheres as condensed matter systems it appears possible that progress can be made on answering these important questions.

\section{Acknowledgments}
I thank Freddy Bouchet, Kerry Emanuel, Grisha Falkovich, Paul Kushner, Paul O'Gorman, Glenn Paquette, Wanming Qi, Florian Sabou, Tapio Schneider, Steve Tobias, Peter Weichman, and Bill Young for helpful discussions.   I would also like to thank Caltech for its hospitality during a sabbatical visit when this article was written, and Jim Langer for reading the article and offering helpful questions and suggestions.  This work was supported in part by NSF grant Nos. DMR-0605619 and CCF-1048701.

\vfill\eject

\section{Mini-Glossary}

The following brief definitions are adapted in part from the Glossary of Meteorology\cite{ams:2000}.

\begin{enumerate}

\item {\bf baroclinic}:  The component of a field that varies with altitude. 

\item {\bf barotropic}: The component of a field that is uniform with altitude.  

\item {\bf direct numerical simulation (DNS)}:  Numerical simulation of equations of motion.  Statistics are accumulated by 
sampling fields during time integration. 

\item {\bf direct statistical simulation (DSS)}:  Numerical simulation of the equations of motion for the statistics themselves.  

\item {\bf eddy}: The departure of a field from its zonal mean.

\item {\bf geostrophic}:  A balance between horizontal Coriolis and pressure forces.  

\item {\bf Hadley cell}: Circulating winds in the tropics that near the surface move toward the equator, rise near the equator, move poleward and then sink back towards the surface near $\pm 30^\circ$ latitude. 

\item {\bf potential temperature}: Temperature that a parcel of dry air would have if brought adiabatically from its initial state to a standard pressure. 

\item {\bf superrotation}:  Flow along the equator in the direction of the planetary rotation. 

\item {\bf zonal mean}: The mean value of a field, averaged over longitude, expressed as a function of latitude. 

\end{enumerate}

\vfill\eject

\section{Important Acronyms}

\begin{enumerate}

\item{\bf CE2}: Second-order cumulant expansion.

\item{\bf CE3}: Third-order cumulant expansion.

\item{\bf DNS}: Direct numerical simulation.

\item{\bf DSS}: Direct statistical simulation.

\item{\bf ESM}: Earth system model.

\item{\bf ME}: Minimum enstrophy.

\item{\bf MEP}: Maximum entropy production

\end{enumerate}

\vfill\eject


 \end{document}